\documentclass[sigconf]{acmart}
\pdfoutput=1

\AtBeginDocument{%
  \providecommand\BibTeX{{%
    \normalfont B\kern-0.5em{\scshape i\kern-0.25em b}\kern-0.8em\TeX}}}

\setcopyright{acmlicensed}

\renewcommand{\textrightarrow}{$\rightarrow$}

\usepackage{epsfig}
\usepackage{graphicx}
\usepackage{amsmath}
\usepackage{amssymb}
\usepackage{array}

\usepackage{xcolor}

\DeclareMathOperator*{\argmin}{arg\,min}

\title{eCNN: A Block-Based and Highly-Parallel CNN Accelerator for Edge Inference}

\begin{document}

\author{Chao-Tsung~Huang} 
\affiliation{%
  \institution{National Tsing Hua University}
  \state{Taiwan, R.O.C.}}

\author{Yu-Chun~Ding} 
\affiliation{%
  \institution{National Tsing Hua University}
  \state{Taiwan, R.O.C.}}
  
\author{Huan-Ching~Wang}
\affiliation{%
  \institution{National Tsing Hua University}
  \state{Taiwan, R.O.C.}}

\author{Chi-Wen~Weng}
\affiliation{%
  \institution{National Tsing Hua University}
  \state{Taiwan, R.O.C.}}

\author{Kai-Ping~Lin}
\affiliation{%
  \institution{National Tsing Hua University}
  \state{Taiwan, R.O.C.}}

\author{Li-Wei~Wang}
\affiliation{%
  \institution{National Tsing Hua University}
  \state{Taiwan, R.O.C.}}

\author{Li-De~Chen}
\affiliation{%
  \institution{National Tsing Hua University}
  \state{Taiwan, R.O.C.}}


\renewcommand{\shortauthors}{C.-T. Huang et al.}

\copyrightyear{2019} 
\acmYear{2019} 
\acmConference[MICRO-52]{The 52nd Annual IEEE/ACM International Symposium on Microarchitecture}{October 12--16, 2019}{Columbus, OH, USA}
\acmBooktitle{The 52nd Annual IEEE/ACM International Symposium on Microarchitecture (MICRO-52), October 12--16, 2019, Columbus, OH, USA}
\acmPrice{15.00}
\acmDOI{10.1145/3352460.3358263}
\acmISBN{978-1-4503-6938-1/19/10}


\begin{abstract}
Convolutional neural networks (CNNs) have recently demonstrated superior quality for computational imaging applications.
Therefore, they have great potential to revolutionize the image pipelines on cameras and displays.
However, it is difficult for conventional CNN accelerators to support ultra-high-resolution videos at the edge due to their considerable DRAM bandwidth and power consumption.
Therefore, finding a further memory- and computation-efficient microarchitecture is crucial to speed up this coming revolution.

In this paper, we approach this goal by considering the inference flow, network model, instruction set, and processor design jointly to optimize hardware performance and image quality.
We apply a block-based inference flow which can eliminate all the DRAM bandwidth for feature maps and accordingly propose a hardware-oriented network model, ERNet, to optimize image quality based on hardware constraints.
Then we devise a coarse-grained instruction set architecture, FBISA, to support power-hungry convolution by massive parallelism.
Finally, we implement an embedded processor---eCNN---which accommodates to ERNet and FBISA with a flexible processing architecture.
Layout results show that it can support high-quality ERNets for super-resolution and denoising at up to 4K Ultra-HD 30 fps while using only DDR-400 and consuming 6.94W on average.
By comparison, the state-of-the-art Diffy uses dual-channel DDR3-2133 and consumes 54.3W to support lower-quality VDSR at Full HD 30 fps.
Lastly, we will also present application examples of high-performance style transfer and object recognition to demonstrate the flexibility of eCNN.
\end{abstract}



\maketitle

\section{Introduction}

Convolutional neural networks (CNNs) recently draw a lot of attention for their great success in the fields of computer vision and computational imaging.
Their hardware accelerators also become an emerging need to enable edge applications.
The performance of pixel throughput and inference quality is determined jointly by model structure, processor architecture, and inference flow.

The CNN model structure has evolved mainly for object recognition, e.g.~from shallow AlexNet \cite{alexnet_2012} to deep \mbox{VGGNet} with small filters \cite{vggnet_2015} and ResNet with residual connections \cite{resnet_2016}.
Several hardware-oriented variants, like depth-wise convolution \cite{mobilenet_2017} and feature squeezing \cite{squeezenet_2017,mobilenetv2_2018}, were also proposed to reduce model complexity for edge inference.
On the other hand, CNNs have also shown dominant performance for computational imaging applications \cite{intro_ipcnn_2017}, such as
image denoising \cite{joint_dm_dn_2016,DnCNN_2017,FFDNet_2018},
super-resolution \cite{SRCNN_2014,VDSR_2016,SRResNet_2017,EDSR_2017},
image deblurring \cite{blind_deblur_2016,dynamic_deblur_2017},
and view synthesis \cite{deep3D_2016,lfsyn_2016}.
They even can provide novel applications which are hard to achieve using traditional methods, like style transfer \cite{st_sr_2016,cyclegan_2017},
DSLR-quality conversion \cite{dslr_2017}, and algorithm mimicking \cite{ipcopy_2017}.
However, there were seldom discussions on hardware-oriented models for computational imaging despite their potential to enable next-generation image pipelines on edge devices.

On the other hand, several hardware accelerators have been proposed for deep neural networks.
For example, \mbox{DaDianNao} \cite{dadiannao_2014}, Cambricon-ACC \cite{cambricon_2016}, TPU \cite{tpu_2017}, and DNPU \cite{dnpu_2018} were designed for general-purpose inference.
In contrast, ShiDianNao \cite{shidiannao_2015}, Eyeriss \cite{eyeriss_2016} , and Morph \cite{morph_2018} were dedicatedly optimized for classification CNNs.
The weight sparsity in these CNNs has been used to reduce computation complexity \cite{cambriconX_2016,cambriconS_2018,cnvlutin_2016,scnn_2017}, and the bit sparsity in activations was deployed in \cite{pragmatic_2017}.
Another approach for saving complexity is to use low-precision computation, such as dynamic fixed-point format  \cite{deephi_2016,precisionscalable_2016} and even binary networks \cite{yodann_2018}.
However, these accelerators are not optimized for computational imaging and also not for high-resolution videos, especially in terms of DRAM bandwidth and computing capability.
Recently, Diffy \cite{diffy_2018} attacked this problem by utilizing the bit sparsity in activation differences to reduce DRAM access and computing power.
But many Diffy tiles and high-end DRAM settings are still required for Full-HD videos.

Finally, the inference flow of a given CNN model determines the data reuse scheme for an accelerator and thus its memory access efficiency.
A systematic approach to partition CNNs into computation sub-blocks was introduced in \cite{blocking_2016}, and several energy-efficient dataflows were analyzed in \cite{eyeriss_2016}.
In particular, a line-based flow was considered for layer fusion in \cite{fusedlayer_2016} which avoids external traffics for feature maps by applying pyramid inference on moving blocks.
For the overlapped features between blocks, a reuse scheme was chosen for fusing up to five CNN layers.
However, the line buffer size will increase linearly with model depth, image width, and channel number.
For example, 9.3MB of SRAM will be required for supporting VDSR \cite{VDSR_2016} in Full HD resolution.
Similar trade-offs between the on-chip SRAM size and off-chip DRAM bandwidth have also been widely studied for image processing applications, such as motion estimation \cite{me_datareuse_2002} and discrete wavelet transform \cite{dwt_datareuse_2005}.

In this work, we aim to enable high-quality edge inference at up to 4K Ultra-HD (UHD) 30 fps for computational imaging tasks.
In particular, we target low-end DRAM settings for cost-effective and power-efficient integration on embedded devices.
We found this challenging goal is hard to achieve by directly accelerating state-of-the-art models which mostly have wide features and deep layers.
Instead, we expand our design space to consider the inference flow, model structure, instruction set, and processor design jointly for optimizing both hardware performance and image quality.

We first propose a block-based truncated-pyramid inference flow which can eliminate all the DRAM bandwidth for feature maps by storing them in on-chip block buffers.
To avoid huge on-chip storage, we choose to recompute the overlapped results between neighboring blocks.
The block buffer size is proportional to model width, and the recomputation overhead almost increases quadratically with model depth.
As a result, these two factors defy the rule of thumb that simply adds more features and more layers to enhance model quality.
Instead, we propose a novel ERNet model to optimize CNNs under these hardware constraints.
Then we construct a feature-block instruction set architecture (FBISA) to support highly-parallel convolution.
It specifies block-buffer-level operations in the fashion of Single Instruction, Multiple Data (SIMD).
In addition, it provides flexibility for programmers and compilers to optimize the computing flow based on different constraints.
Finally, we implement an embedded CNN processor---eCNN---which flexibly accommodates to ERNet and FBISA with highly-parallel filters and locally-distributed parameters.

In summary, the main contributions and findings of this work are:
\begin{itemize}
\item We propose a block-based flow to enable high-resolution inference with low DRAM bandwidth and also analyze its computation and bandwidth overheads. (Section \ref{sec:blockflow})
\item We propose a hardware-aware ERNet to optimize image quality based on hardware constraints and also build training procedures for model optimization and dynamic fixed-point precision.
(Section \ref{sec:ERNet})
\item We devise a coarse-grained FBISA with parallel parameter bitstreams to provide massive computation parallelism efficiently and flexibly.
(Section \ref{sec:FBISA})
\item We design an embedded processor, eCNN, to support FBISA with highly-parallel convolution using 81,920 multipliers.
(Section \ref{sec:eCNN})
\item We train ERNets for image super-resolution (SR) and denoising with 8-bit precision.
In particular, the quality for four-times SR can outperform VDSR \cite{VDSR_2016} by 0.57 dB and 0.44 dB in PSNR when eCNN delivers Full HD and 4K UHD 30 fps respectively.
(Section \ref{ssec:testmodels})
\item Layout results show that eCNN can achieve 41~TOPS (tera operations per second) on 40~nm technology.
It supports high-quality ERNets at up to 4K UHD 30 fps while consuming 6.94W and using DDR-400.
By comparison, Diffy consumes 54.3W and uses dual-channel DDR3-2133 for VDSR at Full HD 30 fps.
(Section \ref{ssec:eCNNperf})
\item Computer-vision tasks can also be well supported by FBISA-compatible models, such as style transfer and object recognition.
(Section \ref{ssec:otherexample})
\end{itemize}

\section{Motivation}
\label{sec:motivation}

Recent research on CNN accelerators mainly focuses on object recognition/detection networks.
Therefore, two specific features for computational imaging networks are not considered for optimization:
1) the spatial resolution of feature maps is not aggressively downsampled and
2) the models are not very sparse.
The former results in a dramatically-high amount of memory bandwidth, and the latter introduces an extremely-high demand of computing power.

\begin{figure}
\centering
  \begin{minipage}[c]{0.45\linewidth}
    \centering
    \includegraphics[height=1.9cm]{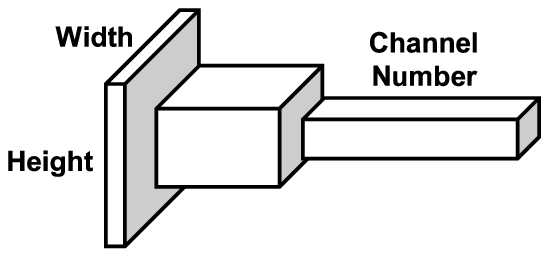} \\
    (a)
  \end{minipage}
  \hfill
  \begin{minipage}[c]{0.45\linewidth}
    \centering
    \includegraphics[height=1.9cm]{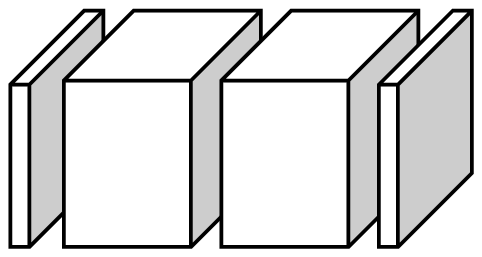} \\
    (b)
  \end{minipage}
\caption{CNN examples in per-layer feature maps for (a) object recognition and (b) computational imaging.
}
\label{fig:fig_model_exp}
\end{figure}

The aggressive downsampling for object recognition is as shown in Fig. \ref{fig:fig_model_exp}(a).
It can extract high-level features and also reduce the data amount for feature maps (volume of cuboids) in deeper layers.
Most of conventional accelerators can thus apply a frame-based inference flow to perform convolution layer-by-layer with limited DRAM bandwidth.
However, this flow will induce a huge amount of DRAM bandwidth for computational imaging networks.
It is because high-resolution feature maps are required to generate texture details and only mild downsampling is allowed \cite{intro_ipcnn_2017}.
Take the plain network without downsampling in Fig. \ref{fig:fig_model_exp}(b) as an example, its corresponding DRAM bandwidth for feature maps, except input and output images, can be derived as
\begin{align}
H \times W \times C \times (D-1) \times fR \times L \times 2, \label{equ:bandwidth}
\end{align}
where $H$ stands for image height, $W$ for image width, $C$ for the number of feature channels (model width), $D$ for model depth, $fR$ for frame rate, $L$ for the bit length of each feature, and the factor $2$ for writing per-layer feature maps into DRAM and then loading back for the next layer.
Accordingly, the 20-layer 64-channel VDSR will require 303 GB/s of memory bandwidth for Full HD 30 fps when using 16-bit features.
Even with the state-of-the-art compression, Diffy still requires dual-channel DDR3-2133 (34 GB/s) to meet this Full-HD specification.
When the resolution is raised to 4K UHD, the DRAM bandwidth will be four times larger and thus unaffordable for small-form-factor and power-limited edge devices.

\begin{figure}
\centering
  \begin{minipage}[c]{0.495\linewidth}
    \centering
    \includegraphics[height=3.4cm]{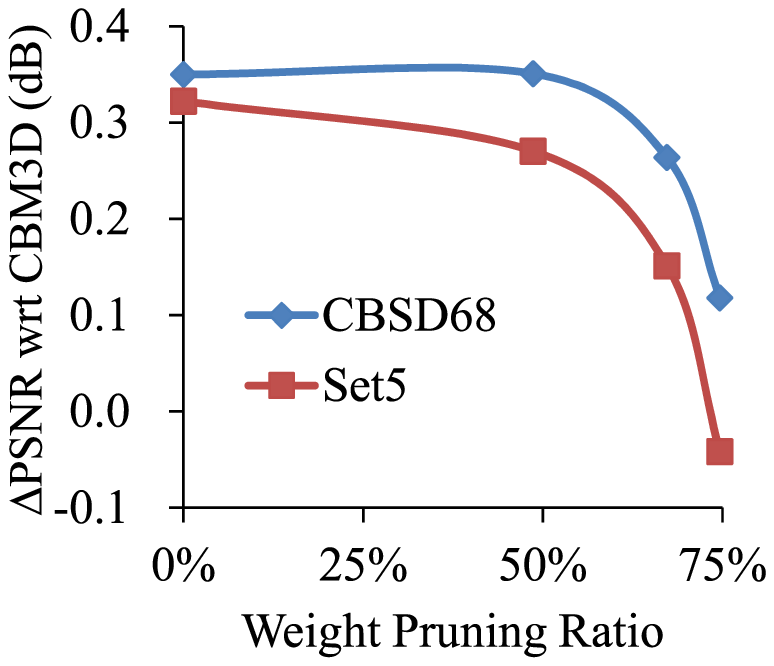} \\
    (a)
  \end{minipage}
  \hfill
  \begin{minipage}[c]{0.495\linewidth}
    \centering
    \includegraphics[height=3.4cm]{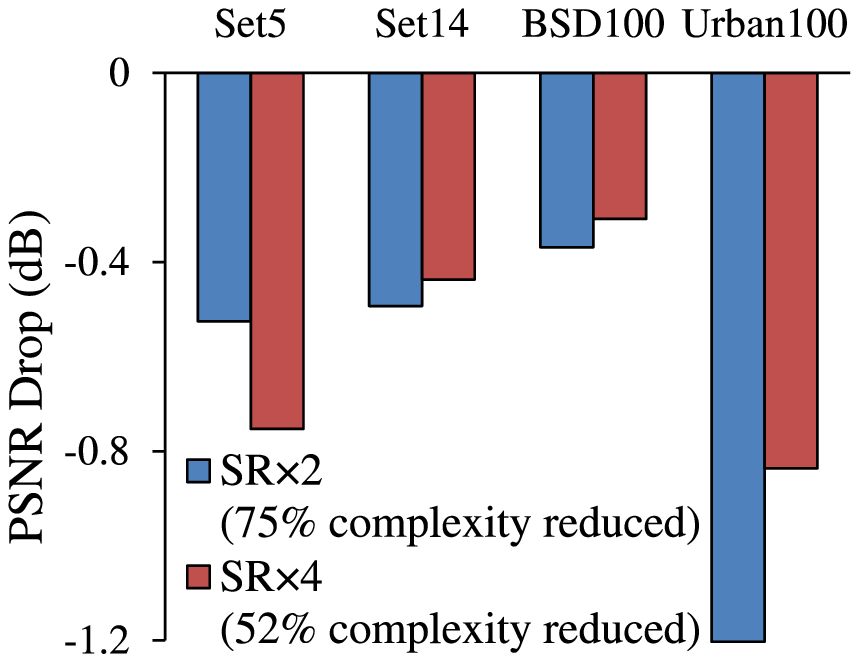} \\
    (b)
  \end{minipage}
\caption{Quality degradation of computational imaging networks for using sparsity techniques.
(a) Weight pruning for DnERNet-B16R1N0.
(b) Depth-wise convolution in residual blocks for EDSR-baseline (SR$\times$2 and SR$\times$4).
}
\label{fig:fig_degradation}
\end{figure}

On the other hand, the sparsity of object recognition models has been deployed to develop many complexity-saving techniques, such as weight pruning \cite{deepcompression_2016} and depth-wise convolution \cite{mobilenet_2017}.
However, computational imaging networks rely on the variety of parameters to extract local features and generate fine textures.
Their image quality is highly related to the model size, so the sparsity techniques could result in significant degradation.
Two such examples are shown in Fig. \ref{fig:fig_degradation}.
One is pruning weights for a denoising DnERNet model (Section \ref{sec:evaluation}).
When pruning 75\% of weights away, its PSNR gain over the benchmark CBM3D \cite{bm3d_2007} drops by 0.2-0.4 dB for two datasets (CBSD68 \cite{cbsd68} and Set5 \cite{set5}) and could even become negative.
Another example is using depth-wise convolution for EDSR-baseline models \cite{EDSR_2017}.
Although 52-75\% of complexity can be saved, the quality drop is 0.3-1.2 dB for four datasets (Set5, Set14 \cite{set14}, BSD100 \cite{cbsd68}, and Urban100 \cite{urban100}) and thus makes the saving unjustified.
Therefore, we need to confront the computation demand for computational imaging CNNs.
Furthermore, high-resolution image generation will make this demand more challenging.
For example, VDSR already demands as high as 83 TOPS for Full HD real-time applications and will require 332 TOPS for 4K UHD.

\begin{figure}
\centering
\includegraphics[width=8.4cm]{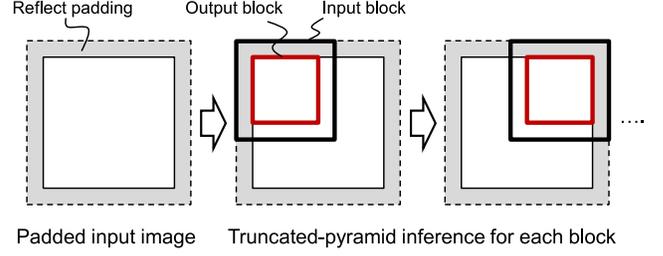} 
\caption{Proposed block-based inference flow.
}
\label{fig:fig_block_flow}
\end{figure}

The issues of huge DRAM bandwidth and computing power motivate us to find a novel approach for ultra-high-resolution CNN acceleration.
In the following, we will propose the block-based inference flow and the hardware-oriented ERNet to resolve the memory issue.
And the computation issue will be addressed by the coarse-grained FBISA and the corresponding highly-parallel eCNN processor.

\section{Block-Based Inference Flow}
\label{sec:blockflow}

The proposed block-based flow is shown in Fig. \ref{fig:fig_block_flow}.
An input image is partitioned into several blocks which can be processed independently, and all the output blocks are then stitched to form a final output image.
In contrast to layer fusion \cite{fusedlayer_2016}, we recompute block-overlapped features to avoid huge on-chip SRAM.
However, this induces additional bandwidth and computation.
To reduce these overheads, we then propose the truncated-pyramid inference (output block larger than one pixel) as detailed in the following.

\begin{figure}
\centering
\includegraphics[width=8.4cm]{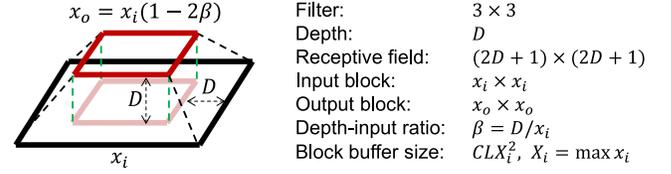} 
\caption{Truncated-pyramid inference for a plain CNN.
}
\label{fig:fig_truc_pyramid}
\end{figure}

To analyze the overheads of this inference flow, we use the plain network in Fig. \ref{fig:fig_truc_pyramid} as an example.
It consists of only CONV3$\times$3 layers, and the receptive field is thus linked directly to the depth $D$.
As the convolution goes to deeper (upper) layers, the effective region will become smaller as a truncated pyramid.
For example, an $x_i\times x_i$ input block will generate an $x_o\times x_o$ output block where $x_o=x_i-2D$.
When the depth (receptive field) is increased, more input blocks and thus more DRAM bandwidth will be required because fewer output pixels are generated for the same input block size.
This bandwidth overhead can be evaluated by a normalized bandwidth ratio (NBR) which is equal to the bandwidth for all input and output blocks over that for an output image:
\begin{align}
NBR = \frac{3x_o^2+3x_i^2}{3x_o^2}=1+\frac{1}{(1-2\beta)^2}, \label{equ:nbr}
\end{align}
where RGB images are considered and $\beta$ is a depth-input ratio, $D/x_i$.
Similarly, there are also computation overheads for the recomputed features among neighboring blocks.
It can be evaluated by a normalized computation ratio (NCR) which represents the computation complexity of this block-based flow over that of the frame-based one (intrinsic):
\begin{align}
NCR = \frac{\mbox{volume of truncated pyramid}}{\mbox{volume of center cuboid}}=\frac{1}{3}+\frac{2}{3}\frac{1-\beta}{(1-2\beta)^2}, \label{equ:ncr}
\end{align}
where the volume (in Fig. \ref{fig:fig_truc_pyramid}) is proportional to the amount of features and therefore that of computing operations.

\begin{figure}[t]
\centering
  \begin{minipage}[c]{0.495\linewidth}
    \centering
    \includegraphics[height=3.4cm]{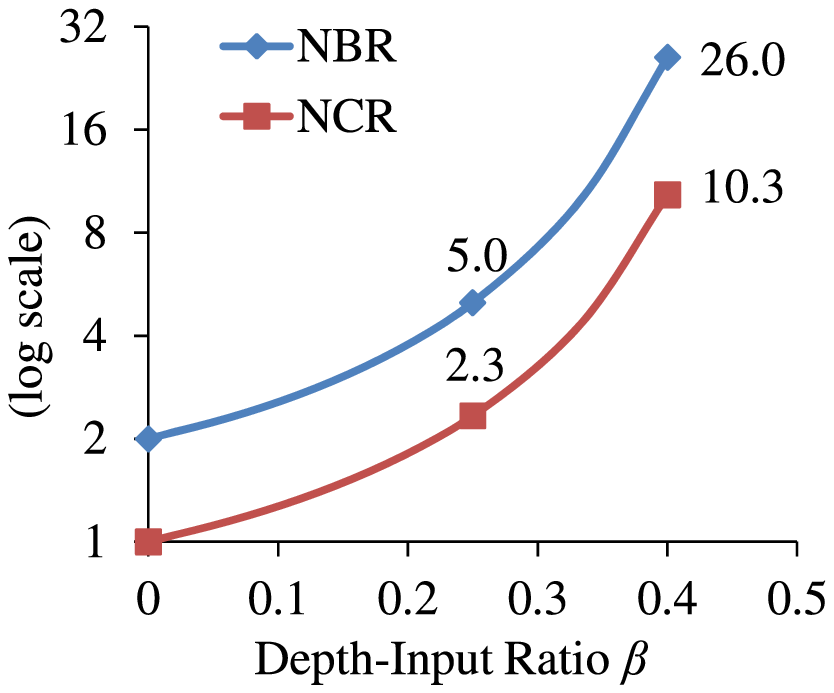} \\
    (a)
  \end{minipage}
  \hfill
  \begin{minipage}[c]{0.495\linewidth}
    \centering
    \includegraphics[height=3.4cm]{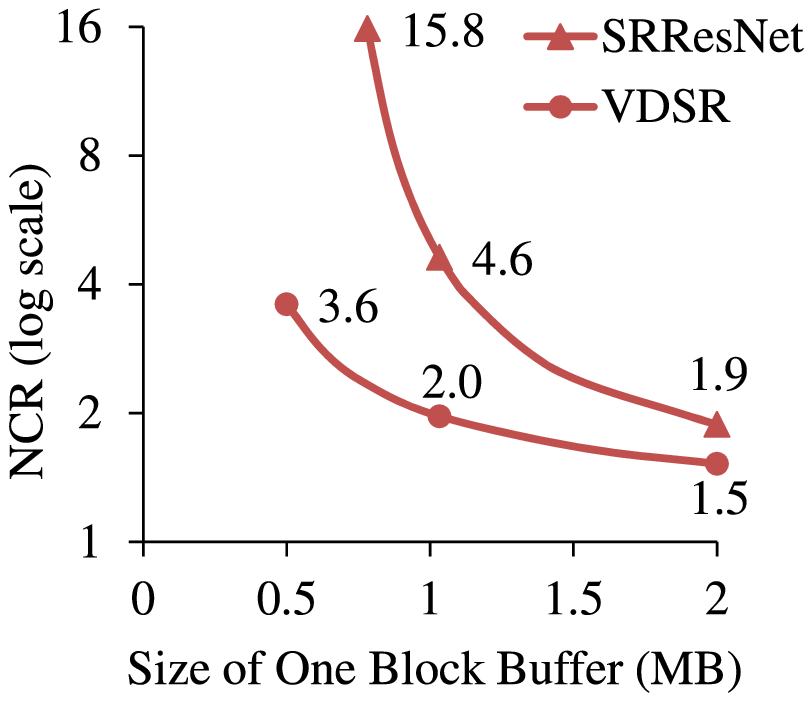} \\
    (b)
  \end{minipage}
\caption{Bandwidth and computation overheads for truncated-pyramid inference.
(a) NBR and NCR versus depth-input ratio for the plain network.
(b) NCR versus block buffer size for 20-layer VDSR and 37-layer SRResNet. ($L=16$)
}
\label{fig:fig_normalized_ratio}
\end{figure}

\begin{figure*}[t]
\centering
  \begin{minipage}[c]{0.465\linewidth}
    \centering
    \includegraphics[height=1.75cm]{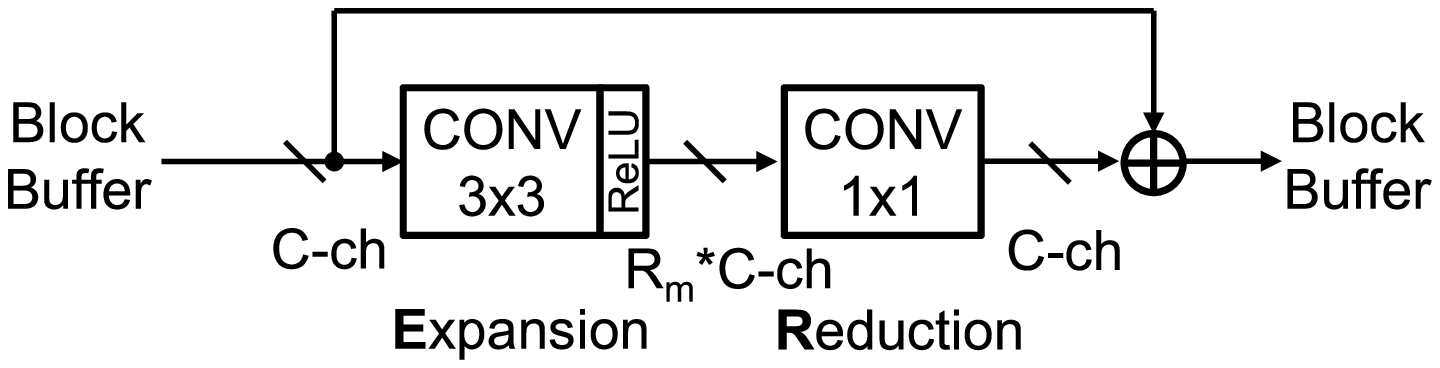} \\
    (a)
  \end{minipage}
  \hfill
  \begin{minipage}[c]{0.525\linewidth}
    \centering
    \includegraphics[height=1.75cm]{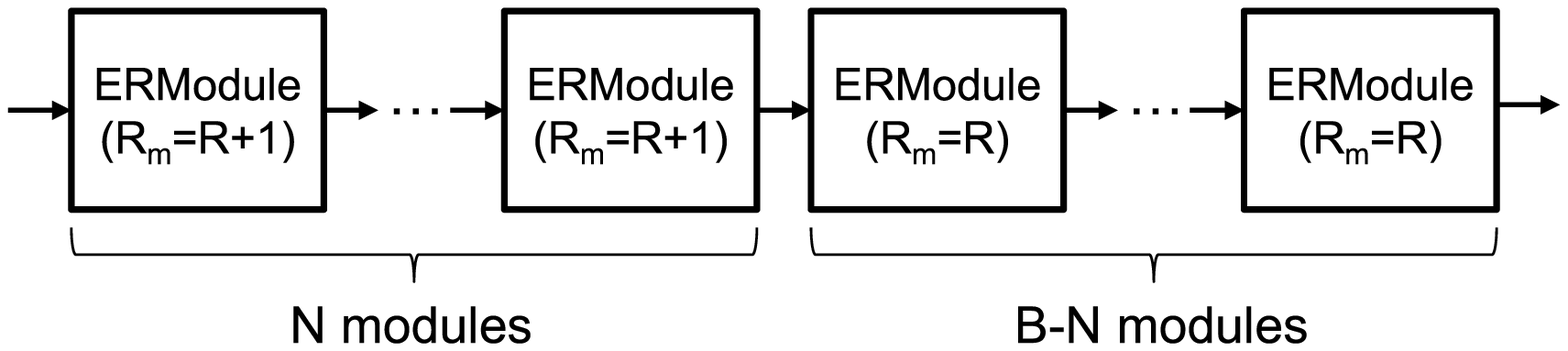} \\
    (b)
  \end{minipage}
\caption{Basic building modules of ERNet: (a) ERModule and (b) connected ERModules ($R_E=R\frac{N}{B}$).
}
\label{fig:fig_ernet}
\end{figure*}

\begin{figure*}
\centering
\includegraphics[width=14cm]{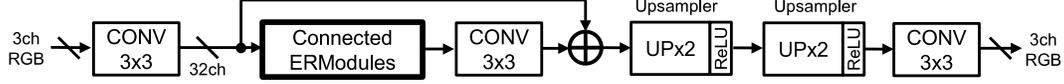}
\caption{SR4ERNet for four-times SR.
}
\label{fig:fig_srx4ernet}
\end{figure*}

These two ratios both grow rapidly with respect to the depth-input ratio $\beta$ as shown in Fig. \ref{fig:fig_normalized_ratio}(a), and they eventually go to infinity when $\beta=0.5$ for $x_o=0$, i.e.~no valid output pixels.
If $\beta$ is not too close to $0.5$, the induced bandwidth overhead is generally acceptable compared to the bandwidth we save.
For example, the NBR is $26\times$ for a large $\beta=0.4$ while the bandwidth overhead of the frame-based flow can be derived as $\frac{2C (D-1)}{3}$ based on (\ref{equ:bandwidth}) and is as high as $811\times$ for VDSR.
However, the computation overhead does become the main side effect of the block-based flow.
We will spend 90\% of the computing power for feature recomputation when $\beta$ approaches $0.4$.
To avoid this situation, we will need to adopt larger block buffers to reduce $\beta$ for deeper networks.

The evaluation of the above plain example can be extended to more complicated models, and the conclusions on bandwidth and computation overheads are similar.
Most of state-of-the-art networks have deep layers of 3$\times$3 (or larger) filters and wide channels of feature maps, e.g. $C\geq 64$.
Therefore, they will either require huge block buffers of size $C L X_i^2$ to reduce NCR or suffer significant computation overheads for using small buffers to save area.
Note that usually more than one block buffer will be required for switching between input and output layers, which makes the area cost more severe.

Fig. \ref{fig:fig_normalized_ratio}(b) shows this trade-off between the NCR and block buffer size for VDSR and also a state-of-the-art SRResNet \cite{SRResNet_2017} which outperforms VDSR by 0.6 dB \cite{EDSR_2017}.
The NCR for the 20-layer VDSR is well controlled as $2\times$ using 1MB block buffers.
But the 37-layer SRResNet needs around 2MB to have a similar NCR.
Using smaller block buffers to save area for SRResNet will make the NCR skyrocket quickly.
Therefore, with the block-based flow it is difficult to have a low NCR and use small buffers simultaneously for deep high-quality networks.
In the following, we will achieve this goal by considering these hardware constraints as early as model construction and accordingly introduce the ERNet.

\section{ERN\lowercase{et}}
\label{sec:ERNet}

We will first introduce the basic building modules of ERNet and then present a procedure for model optimization.
The quantization method we adopt for dynamic fixed-point precision will also be discussed.

\subsection{Model Structure}
\label{ssec:ERModule}

Consider a thin network which can use small block buffers.
To increase its capacity without enlarging the NCR and buffer area, we explore another direction of model construction by temporarily expanding the model width.
This is achieved by the ERModule shown in Fig. \ref{fig:fig_ernet}(a).
It uses a CONV3$\times$3 layer to expand the model width by $R_m$ times and a following CONV1$\times$1 to reduce it back.
A residual connection is added for robust training.
All the operations are performed internally without accessing to block buffers.
Therefore, we can pump complexity into ERModule to improve image quality with the same block buffer size and model depth.

We only consider integer expansion ratios for $R_m$ to guarantee high hardware utilization.
To increase model flexibility, we further construct a larger building block by connecting $B$ ERModules as shown in Fig. \ref{fig:fig_ernet}(b).
The first $N$ modules can be assigned an incremented $R_m=R+1$ to make the overall expansion ratio $R_E$ as a fraction $R\frac{N}{B}$.
Accordingly, we now have two model hyperparameters to build networks: $B$ for increasing depth and $R_E$ for pumping complexity.

Fig. \ref{fig:fig_srx4ernet} shows a model example, SR4ERNet, for performing four-times SR.
It basically replaces the residual blocks in SRResNet \cite{SRResNet_2017} or EDSR-baseline \cite{EDSR_2017} by ERModules.
It starts with small images of 1/4 size in width and height and uses two pixel-shuffle upsamplers to restore full resolution.
In addition, we reduce the channel number from 64 to 32 for saving area for block buffers.

\subsection{Model Optimization}
\label{ssec:hyper}

\begin{figure}
\centering
   \includegraphics[width=8.2cm]{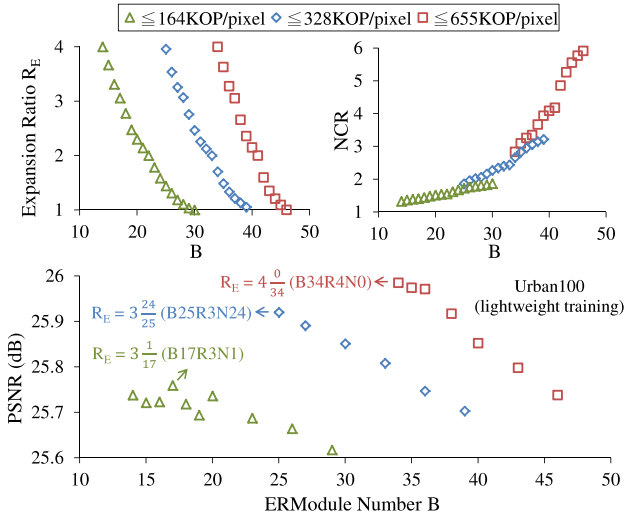} \\
\caption{Model scanning of SR4ERNet for three computation constraints with $x_i=128$.
}
\label{fig:fig_modelsel_srx4}
\end{figure}

The major hardware constraint considered here is the overall computation complexity, i.e.~NCR$\times$(intrinsic complexity), since the bandwidth overhead is usually small.
Each complexity target will correspond to a real-time throughput, and we aim to optimize image quality under such constraints.
Our model selection procedure can be illustrated using SR4ERNet as shown in Fig. \ref{fig:fig_modelsel_srx4}.
We assume the size of input blocks is 128$\times$128 and consider three computation constraints: 164, 328, and 655 KOP/pixel (thousand operations per output pixel). 

First of all, we derive the largest possible expansion ratio $R_E$ for each module number $B$ under each constraint, and we choose $R_E \leq 4$ as a system upper bound.
As shown at the top of Fig. \ref{fig:fig_modelsel_srx4}, $R_E$ will decrease quickly as the model depth grows with $B$ because of the fast-increasing NCR.
In the case of 655 KOP/pixel, NCR can be as high as 2.8-5.9$\times$, and the corresponding intrinsic complexity is as low as 223-107 KOP/pixel.
Note that deeper networks do not necessarily perform better now because of their lower intrinsic complexity.
Then we scan all of these candidate models with a lightweight training setting, e.g.~using smaller patches and fewer mini-batches.
After that, we test their image quality using validation datasets and pick the best model for each constraint as shown at the bottom of Fig. \ref{fig:fig_modelsel_srx4}.
Finally, we will further polish the best models by retraining them with a full setting.
In this example, the highest-quality SR4ERNet-B34R4N0 can even outperform SRResNet by 0.04 dB using a thinner and less-complex network.

\subsection{Dynamic Fixed-Point Precision}
\label{ssec:dfp}

\begin{figure}
\centering
   \includegraphics[width=8.3cm]{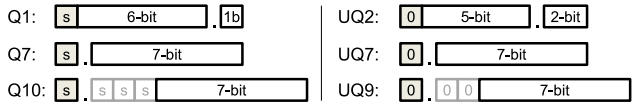} \\
\caption{Examples of 8-bit Q-formats: Q$n$ and UQ$n$.
}
\label{fig:fig_qformat}
\end{figure}

We further quantize the polished models for saving computation logics and on-chip memory.
The multiplications and block buffers are both considered in 8-bit precision while the internal partial sums are accumulated in full precision to preserve quality.
We adopt the fixed-point Q-format for decimals as illustrated in Fig. \ref{fig:fig_qformat}.
Q$n$ and UQ$n$ stand for signed and unsigned values respectively, and $n$ is the fractional position of the last effective bit.
We apply dynamic fixed-point precision to optimize image quality, so each convolution layer has its own Q-formats for weights, biases, and feature outputs, respectively.
We then build a two-stage procedure, quantization and fine-tuning, based on \cite{fgdfp_weight_2015,deephi_2016,ristretto_2016}.

The quantization stage is to determine the best fractional precision $\hat{n}$ of each Q-format.
With a collection $\Omega$ of the corresponding floating-point values, we can use either L1-norm \cite{deephi_2016} or L2-norm \cite{fgdfp_weight_2015} errors for the optimization:
\begin{align}
\hat{n}_l = \argmin_n \sum_{x\in\Omega} | x-Q_n(x)|^l, \:\: l\in\{1,2\}, \label{equ:l2_q}
\end{align}
where the quantization function $Q_n(\cdot)$ performs clipping and rounding for precision $n$.
The value distributions of parameters are directly derived from the floating-point model, and those of feature maps are collected by inferencing on the training dataset.
Since this 8-bit quantization induces up to 3.69 dB of PSNR loss for denoising and SR, we then use the fine-tuning method in \cite{ristretto_2016} to refine these quantized parameters.
For calculating gradients more accurately, we add clipped ReLU (rectified linear unit) functions to the model for the clipping behavior of $Q_n(\cdot)$.
As a result, the fixed-point ERNet has only 0.08 dB of PSNR degradation on average.

\section{FBISA}
\label{sec:FBISA}

We design the SIMD instruction set, FBISA, to support the truncated-pyramid inference for fully convolutional networks.
To increase its flexibility, we also include a zero-padded inference type and a variant of upsamplers and downsamplers.
FBISA provides massive parallelism by coarse-grained instructions between feature blocks and internal accessing of parameter memories which will be introduced sequentially.

\subsection{Instruction Set}
\label{ssec:isa}

\begin{figure}
\centering
   \includegraphics[width=8.3cm]{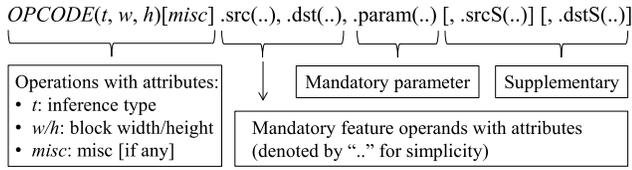} \\
\caption{FBISA instruction format.
}
\label{fig:fig_instruction_fmt}
\end{figure}

Fig. \ref{fig:fig_instruction_fmt} shows the instruction format.
An opcode can specify a convolution task with specific attributes, e.g.~inference type and block size.
There are two kinds of operands for features and parameters, respectively, and they also have their own attributes, in particular for Q-formats.
For the feature operands, there are two mandatory types to inform the source (\emph{src}) and destination (\emph{dst}) of the convolution.
In addition, two supplementary ones (\emph{srcS}/\emph{dstS}) are designed to support feature accumulation among instructions, such as skip/residual connection or partial sums.
Finally, the parameter operand specifies where to access the corresponding weights and biases in parameter memories for the opcode.
For these operands, we use named expressions, instead of the conventional ordered ones, to improve readability.

\begin{table}
\centering
   \includegraphics[width=8.3cm]{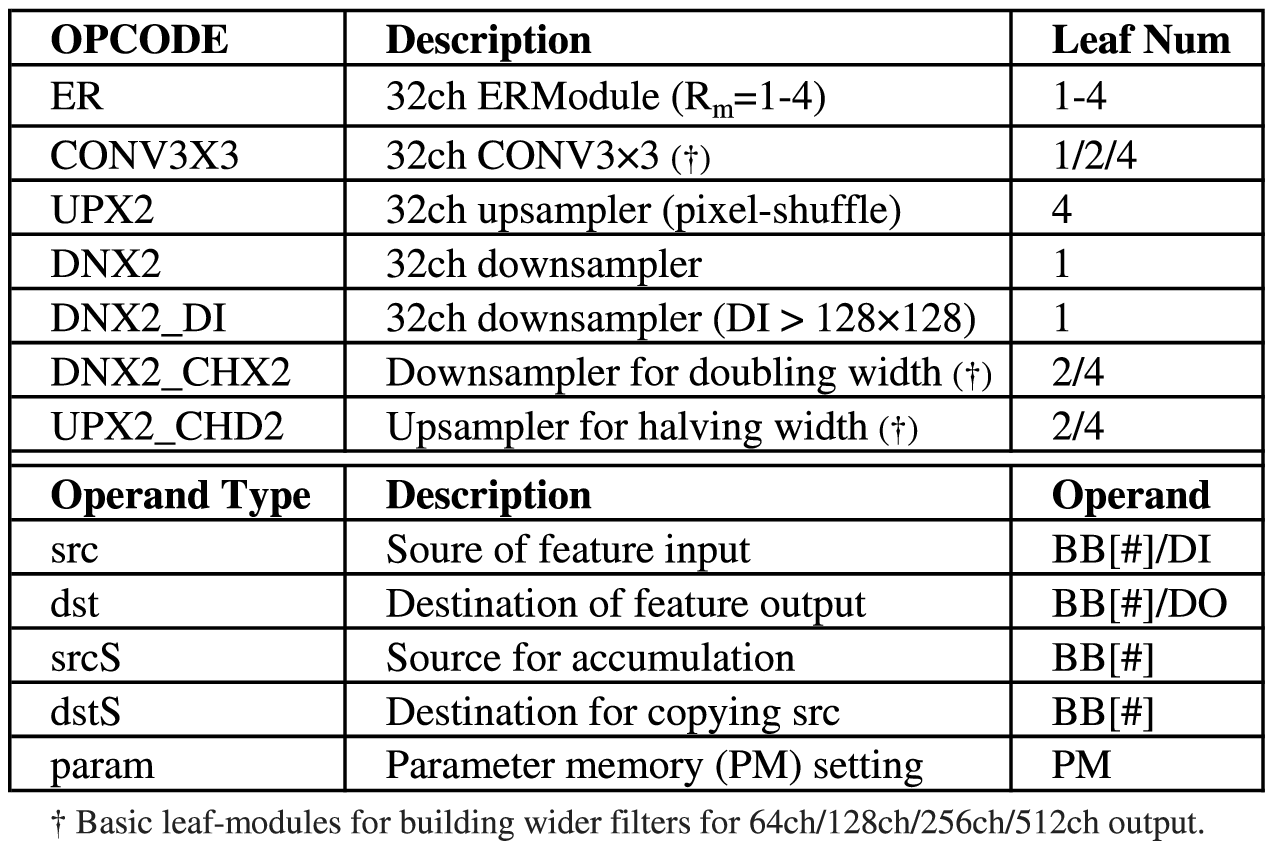} \\
\caption{FBISA instruction overview.
}
\label{tab:tab_instruction_overview}
\end{table}

Table \ref{tab:tab_instruction_overview} provides an overview of the instruction set.
The smallest computing task in FBISA is called a leaf-module, and it performs a 32ch-to-32ch CONV3$\times$3 filter on one feature block.
Each opcode can contain up to four leaf-modules based on its attribute.
The opcodes mainly differ on their usage purposes and thus on the post-processing of outputs.
For example, the opcode \emph{UPX2} shuffles pixels for spatial upsampling while \emph{DNX2} performs strided- or max-pooling for downsampling.
In particular, \emph{ER} is devised specifically for ERModule and its leaf-module has an additional 32ch CONV1$\times$1 for feature reduction.
If wider filters are required for CONV3$\times$3, they can be constructed by using 32ch-based opcodes and accumulating partial sums via \emph{srcS}.

Regarding the feature operands, we apply two strategies to provide efficient data movement for highly-parallel convolution.
First, they are specified on the basis of block buffers (BBs), instead of conventional small registers or vectors.
Therefore, the internal partial sums for one instruction can be accumulated inside hardware-optimized datapaths without accessing to large SRAM or DRAM which is mostly bandwidth-limited and power-hungry.
Another advantage is that we can have small-sized programs and avoid complex compilers.
For example, the high-quality SR4ERNet-B34R4N0 uses only 45 lines of instructions.

The second strategy is \emph{not} using conventional load-store instructions for external feature reading and writing.
Instead, we devise operands \emph{DI} and \emph{DO} as virtual block buffers for data input and output respectively.
They can be implemented by FIFO interfaces and stream data in the same way as normal block buffers.
Therefore, the processor pipeline can be fully optimized for a computation-only instruction set.
This strategy also decouples FBISA from the data structure in the main memory for better system integration portability.

\subsection{Parameter Format}
\label{ssec:ec}

The data movement of parameters is also of paramount importance for CNN acceleration.
To avoid retransmitting parameters for each block, we keep them in internal parameter memories for reuse.
This is feasible thanks to the small-sized computational imaging networks.
For example, the numbers of parameters in VDSR and SRResNet are 651K and 1479K respectively while it is 11M for ResNet-18 \cite{resnet_2016}.

\begin{figure}
\centering
   \includegraphics[width=8.3cm]{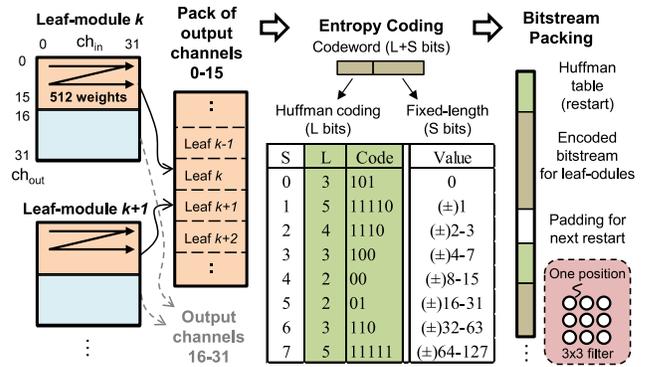} \\
\caption{Weight bitstream for one filter position.
}
\label{fig:fig_bitstream_format}
\end{figure}

In FBISA, we split the filter weights into 20 bistreams to enable parallel loading and distribution of them in the processor: 18 for CONV3$\times$3 and two for CONV1$\times$1.
Each spatial filter position corresponds to two bitstreams for its first and second halves of output channels in leaf-modules.
Fig. \ref{fig:fig_bitstream_format} shows the format of one such bitstream.
The weights are compressed to increase the supported model size, and we adopt the DC Huffman coding in JPEG \cite{jpeg}.
This simple coding algorithm enables fast and parallel decoding with small hardware overheads.
We also found that the weights are mostly uncorrelated, so differential encoding is unnecessary.
On the other hand, the filter biases are gathered in another one bistream and compressed in the same way.

A decoding restart mechanism is further devised to enable parameter reuse between different instructions.
In this case, a byte-aligned address referred to the bias bitstream should be specified as the restart attribute in the parameter operand.
The Huffman table will be placed first and followed by the encoded bitstream.
For the 20 weight bitstreams, their restart addresses will be synchronized to 8$\times$ of the restart attribute.
It is because each of them contains 512 coefficients for one leaf-module but the bias bistream only has 64.
Finally, the 21 bistreams are synchronized for each restart segment by padding shorter ones.
Regarding compression efficiency, we found that one Huffman table for each restart segment in each bitstream is sufficient since the 8-bit quantized parameters have similar distributions.
As a result, the compression ratio is around 1.1-1.5$\times$ for denoising and SR ERNets.

\section{\lowercase{e}CNN}
\label{sec:eCNN}

This embedded processor is implemented to support FBISA with highly-parallel convolution for high-performance and also power-efficient computing.
In the following, we will first present its system architecture for top control and embedded integration.
Then we will introduce its two main functional units for distributing parameters and performing convolution, respectively.

\subsection{System Architecture}
\label{ssec:sys_arch}

\subsubsection{Processing Flow}
\label{sssec:proc_flow}

\begin{figure}
\centering
   \includegraphics[width=8.3cm]{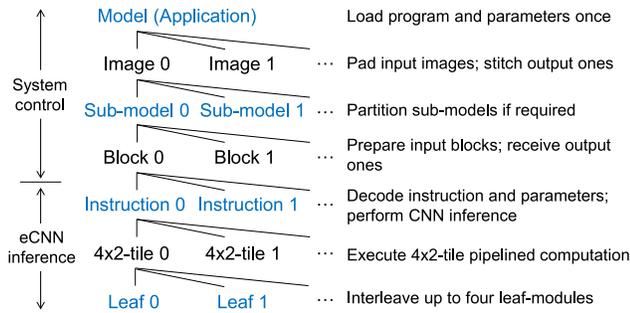} \\
\caption{Processing flow for system control and eCNN.
}
\label{fig:fig_processing_flow}
\end{figure}

For one target model, its program and parameters are loaded into eCNN only once.
Then the inference for each image is performed based on a model hierarchy: sub-model(s), instructions, and leaf-modules.
On the other head, the image is also processed with a pixel-grouping hierarchy: blocks and tiles (4$\times$2 32ch features).
These two hierarchies are interleaved to form a flexible processing flow shown in Fig. \ref{fig:fig_processing_flow}.
In particular, a deep model can be partitioned into few shallower sub-models to reduce computation overheads.
However, the intermediate features between them may increase DRAM bandwidth sharply, which is a performance tradeoff.

\begin{figure*}[t]
\centering
   \includegraphics[width=15.0cm]{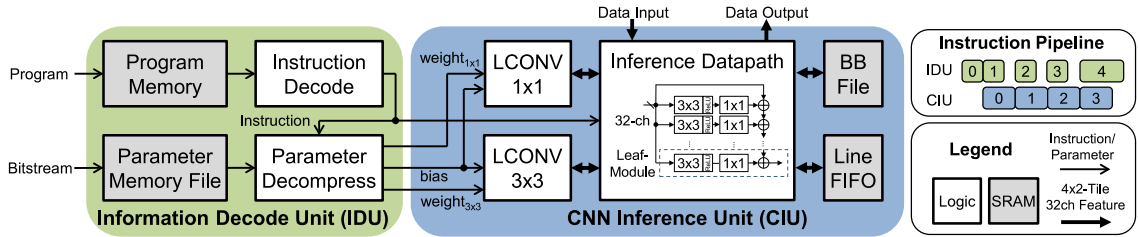} \\
\caption{eCNN system block diagram and instruction pipelining scheme.
}
\label{fig:fig_system_arch}
\end{figure*}

\begin{figure*}[t]
\centering
   \includegraphics[width=15.0cm]{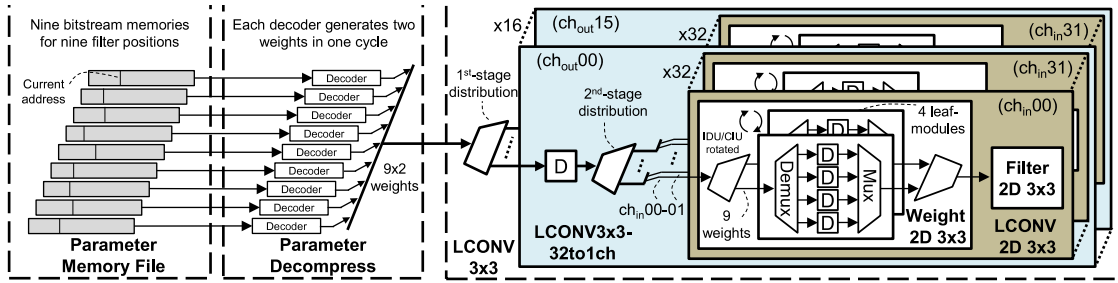} \\
\caption{Weight distribution scheme for the first half of output channels in the LCONV3$\times$3 engine.
}
\label{fig:fig_parameter_dist}
\end{figure*}

The eCNN is designed for embedded integration, e.g.~controlled by a main processor and connected to a DMA controller via FIFO interfaces.
These system transactions can be handled on a block basis without inducing heavy system burdens.
The eCNN then accelerates the computation for each block in an instruction-by-instruction fashion.
For each instruction, it calculates one 32ch leaf-module for a 4$\times$2-tile in one cycle.
And, for each tile, the leaf-modules of this same instruction are calculated consecutively to accumulate their partial sums on-the-fly without precision loss and SRAM access.
After all of the specified 4$\times$2-tiles are processed, the eCNN will repeat similar acceleration for the next instruction.

\subsubsection{Block Diagram}
\label{sssec:sys_diag}

Fig. \ref{fig:fig_system_arch} shows the system block diagram to implement the above-mentioned processing flow.
It consists of two functional units: information decode unit (IDU) and CNN inference unit (CIU).
The IDU is responsible to decode instructions and parameters, and the CIU computes the corresponding convolution.
To enable highly-parallel computing, we deploy a massive amount of multipliers in two convolution engines in CIU: LCONV3$\times$3 and LCONV1$\times$1.
They perform the 32ch CONV3$\times$3 and CONV1$\times$1, respectively, in each leaf-module, and the latter is used for ERModule.

We keep all of the model parameters in IDU to avoid excessive external bandwidth for parameter retransmission.
However, the multipliers need to access up to 10,240 weights for each leaf-module in one single cycle.
The throughput is much higher than the affordable bandwidth of the parameter memories.
Therefore, we devise an instruction pipelining scheme to distribute parameters efficiently.
As a result, the IDU has a whole pipeline stage to progressively decode the parameters for one instruction.
Meanwhile, they are sequentially sent to the locally-distributed registers inside the multipliers of CIU and will be used for the convolution in the next pipeline stage.
In the following, we will introduce the implementation details for the IDU and CIU.

\subsection{Information Decode Unit (IDU)}
\label{ssec:idu}

For each instruction, the IDU will first decode its opcode and operands and then trigger a parameter decompression procedure.
The 21 parameter bitstreams mentioned in Section \ref{ssec:ec} are stored in 21 corresponding memories and decoded by 21 parallel decoders.
For each leaf-module, a weight decoder is responsible to decode 512 weights while the bias one generates at most 64 biases.
All the parameters follow a ping-pong distribution scheme between the IDU and CIU.

Fig. \ref{fig:fig_parameter_dist} shows the distribution scheme for the weights in the first half of output channels in CONV3$\times$3, and the other cases are similar.
Nine decoders are deployed for nine filter positions, and each one decodes two weights in one cycle for two input channels.
The decoded weights are then distributed through a two-stage network: the first and second stages are for output and input channels respectively.
In particular, there are 16$\times$32 local register files (Weight 2D 3$\times$3) accompanied with their corresponding 2D filters (Filter 2D 3$\times$3).
Each register file will keep the decoded parameters in a ping-pong fashion for the CIU convolution in the next instruction pipeline.
Also, it can switch between four leaf-modules for the consecutive computation in one instruction.
In most cases, the IDU decodes one leaf-module in 256 cycles and completes one instruction faster than the CIU of which the run time is proportional to the number of 4$\times$2-tiles.

\subsection{CNN Inference Unit (CIU)}
\label{ssec:ciu}

All of the computation in the CIU goes through an inference datapath which is closely coupled to the two convolution engines and three block buffers (BBs) as shown in Fig. \ref{fig:fig_system_arch}.
The details of these designs are discussed as follows.

\begin{figure*}[t]
\centering
   \includegraphics[width=15cm]{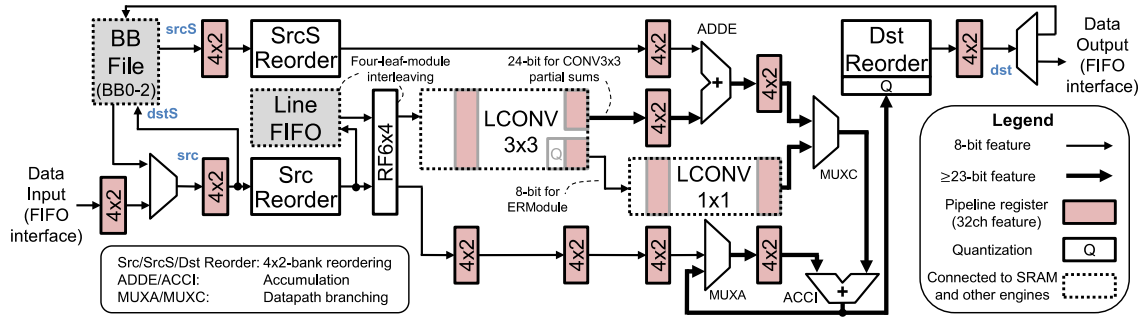} \\
\caption{Inference datapath engine.
}
\label{fig:fig_ecnn_datapath}
\end{figure*}

\begin{figure}
\centering
   \includegraphics[width=8.1cm]{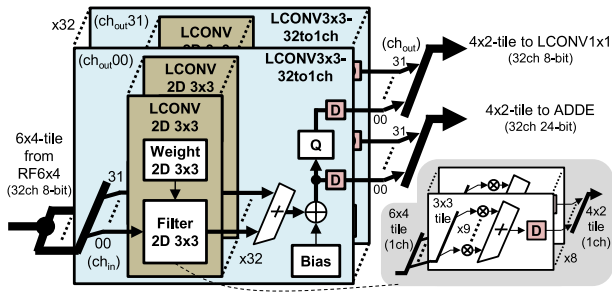} \\
\caption{LCONV3$\times$3 engine.
}
\label{fig:fig_lconv3x3}
\end{figure}

\subsubsection{Tile-Pipelined Inference Datapath Engine}
\label{sssec:datapath}

The highly-parallel convolution is prone to inefficiency of data movement and inflexibility of model supporting.
Thus we carefully designed the inference datapath engine to alleviate these issues.
It follows a 32ch 4$\times$2-tile pipeline as shown in Fig. \ref{fig:fig_ecnn_datapath} and mainly consists of two functions: input preparation for the LCONV3$\times$3 engine and output post-processing for different opcodes and operands.

The first function prepares input 6$\times$4-tiles for 3$\times$3 filtering.
However, a 6$\times$4-tile is three times as large as a 4$\times$2-tile and thus induces heavy bandwidth for block buffers.
To reduce the bandwidth, we only read 4$\times$2-tiles from block buffers and store them in a line FIFO buffer.
Then for each leaf-module the corresponding 6$\times$4-tile can be rearranged in a register file (RF6$\times$4), and up to four leaf-modules are supported.
In addition, a data reordering circuit (Src Reorder) is used to address a tile misalignment issue (Section \ref{sssec:blockbuffermap}).

The second function, output post-processing, provides model flexibility and there are five sub-functions supported:
1) ERModule through the LCONV1$\times$1 engine;
2) Accumulation which uses an adder (ADDE) for calculating cross-instruction partial sums and another adder (ACCI) for internal ones;
3) Upsampling which writes data in pixel-shuffle order (by Dst Reorder);
4) Downsampling for strided- or max-pooling (also by Dst Reorder);
5) Quantization which quantizes features or partial sums to their 8-bit Q-format before going to block buffers or output FIFO interface.
In addition, one 8-bit quantization circuit is used inside LCONV3$\times$3 to reduce the input bitwidth of LCONV1$\times$1 for saving area.

\subsubsection{Highly-Parallel Convolution Engine}
\label{sssec:lconv3x3}

The LCONV3$\times$3 and LCONV1$\times$1 engines serve a 4$\times$2-tile in one cycle for 3$\times$3 and 1$\times$1 filtering respectively.
They employ the weight-stationary strategy \cite{eyeriss_2016} to optimize data reuse for the block-based inference.
Also, compared to the conventional accelerators with much fewer multipliers, their massive parallelism enables power-efficient accumulation of internal partial sums.
It is because their communications are all locally hardwired without going through additional register files, SRAM, or DRAM.
For example, the LCONV3$\times$3 engine contains 32$\times$32 2D filters (Filter 2D 3$\times$3) as shown in Fig. \ref{fig:fig_lconv3x3}.
And each 2D filter shares the same 3$\times$3 weights for a 4$\times$2-tile and reuses them for a block.

\begin{figure}
\centering
  \includegraphics[width=8.3cm]{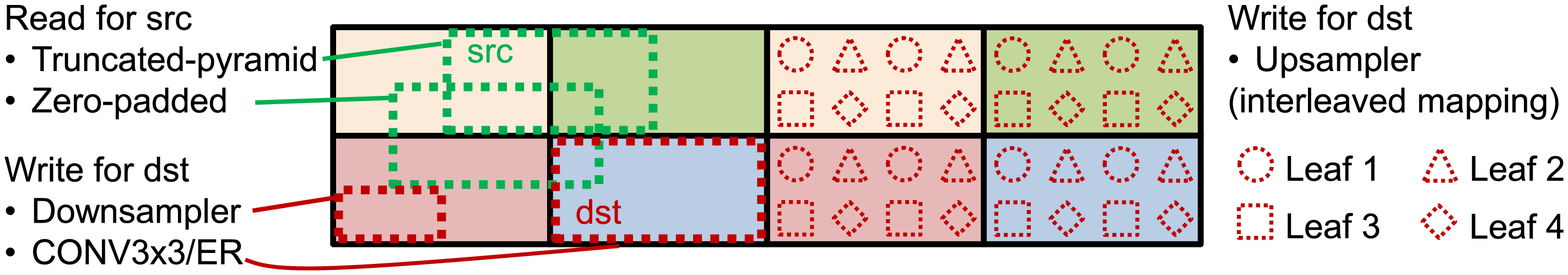} \\
  (a) \\
  \begin{minipage}[c]{0.52\linewidth}
    \centering
    \includegraphics[height=2.05cm]{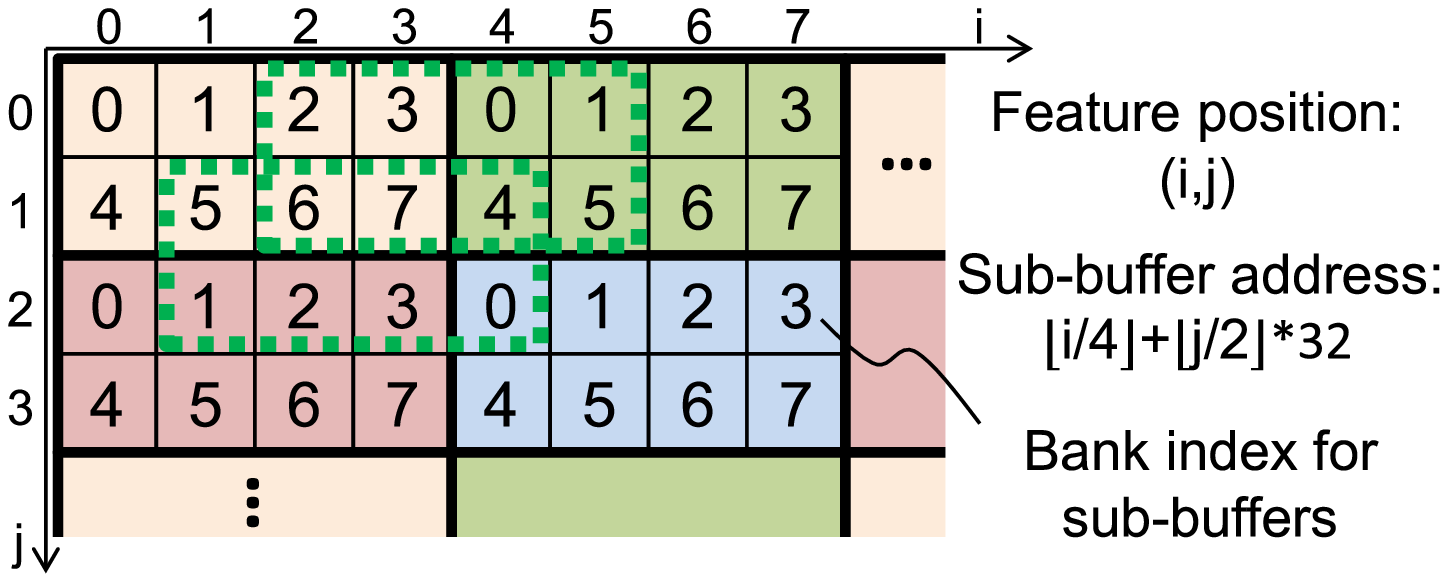} \\
    (b)
  \end{minipage}
  \hfill
  \begin{minipage}[c]{0.38\linewidth}
    \centering
    \includegraphics[height=2.05cm]{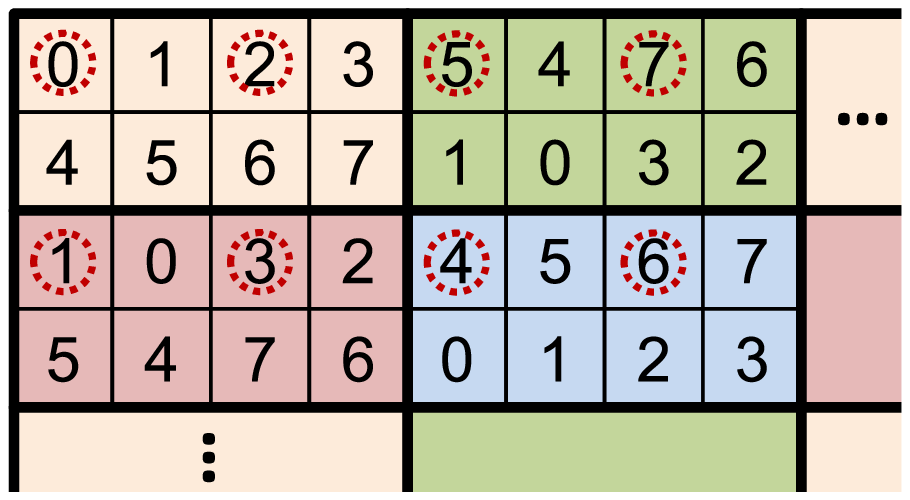} \\
    (c)
  \end{minipage}

\caption{Eight-bank BB implementation. (a) Access patterns for reading and writing 4$\times$2-tiles in a block. (b) Normal bank mapping. (c) Interleaved bank mapping. 
}
\label{fig:fig_bb_map}
\end{figure}

\subsubsection{Eight-Bank Block Buffer Mapping}
\label{sssec:blockbuffermap}

The highly-parallel data movement also brings a misalignment issue for the block buffers:
the features are stored in 4$\times$2-tiles but their accesses are not always tile-aligned.
To address this issue, we implement each block buffer using eight sub-buffer banks as shown in Fig. \ref{fig:fig_bb_map}.
A normal mapping is sufficient for all cases except for pixel-shuffle upsampling which causes sub-buffer conflicts; 
therefore, another interleaved mapping is devised to resolve this issue.

\section{Evaluation}
\label{sec:evaluation}

The configurations of our implementation are listed in Table \ref{tab:tab_eCNN_spec}.
The three computation constraints used for model optimization correspond to three real-time specifications: 4K UHD 30fps (UHD30), Full HD 60fps (HD60), and Full HD 30fps (HD30).
In the following, we will first present the results for ERNet models and then the layout performance for the eCNN processor.
We will also introduce two application examples in computer vision, style transfer and object recognition, to demonstrate the flexibility of our approach.

\begin{table}
\centering
   \includegraphics[width=8.3cm]{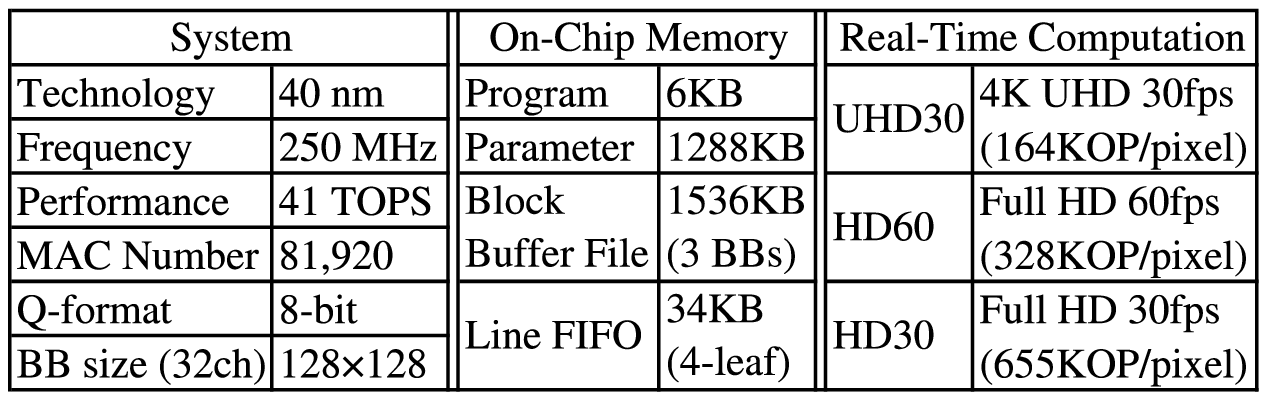} \\
\caption{eCNN configurations.
}
\label{tab:tab_eCNN_spec}
\end{table}

\subsection{ERNet Models}
\label{ssec:testmodels}

\textbf{Model structure.}
In addition to SR4ERNet for four-times SR, we also implement SR2ERNet and DnERNet for two-times SR and denoising respectively.
Their models are derived by accordingly removing one and two upsamplers from the SR4ERNet in Fig. \ref{fig:fig_srx4ernet}.
We do not use batch normalization layers as suggested in \cite{EDSR_2017}.
Also, we pad 29 zero-valued channels for RGB images to form 32ch inputs for eCNN.

\textbf{Training.}
The hyper-parameters are listed in Table \ref{tab:tab_training_hyperparameter} for the three stages of our training procedure: model scanning, polishment, and fine-tuning for quantization.
The scanning uses lightweight settings for speeding up the process, and then the other two apply heavy settings for improving image quality.
We train on two datasets, DIV2K \cite{div2k} and Waterloo Exploration \cite{waterloo}, for the following comparison with the state-of-the-art networks of SR and denoising, respectively.

\begin{table}
\centering
   \includegraphics[width=8.3cm]{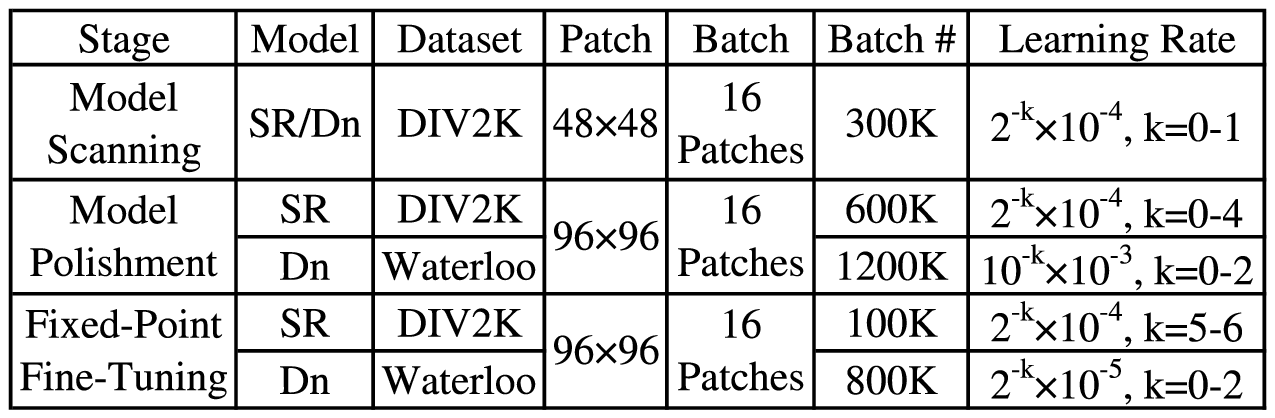} \\
\caption{ERNet training settings.
}
\label{tab:tab_training_hyperparameter}
\end{table}

\textbf{Polished models.}
The PSNR performance of the picked models and their polished results is shown in Table \ref{tab:tab_model_optimization}.
For comparison, we include VDSR and SRResNet (implementation in \cite{EDSR_2017}) for SR and also CBM3D and FFDNet \cite{FFDNet_2018} for denoising.
For the HD30 specification, the hardware-constrained ERNet models can achieve similar quality compared to the state-of-the-art SRResNet and FFDNet.
When we increase the specification, the PSNR performance will drop as the intrinsic complexity goes down.
However, for UHD30 the SR4ERNet can still outperform VDSR by 0.49 dB while the SR2ERNet and DnERNet are comparable to the benchmark VDSR and CBM3D respectively.

\begin{table}
\centering
   \includegraphics[width=8.3cm]{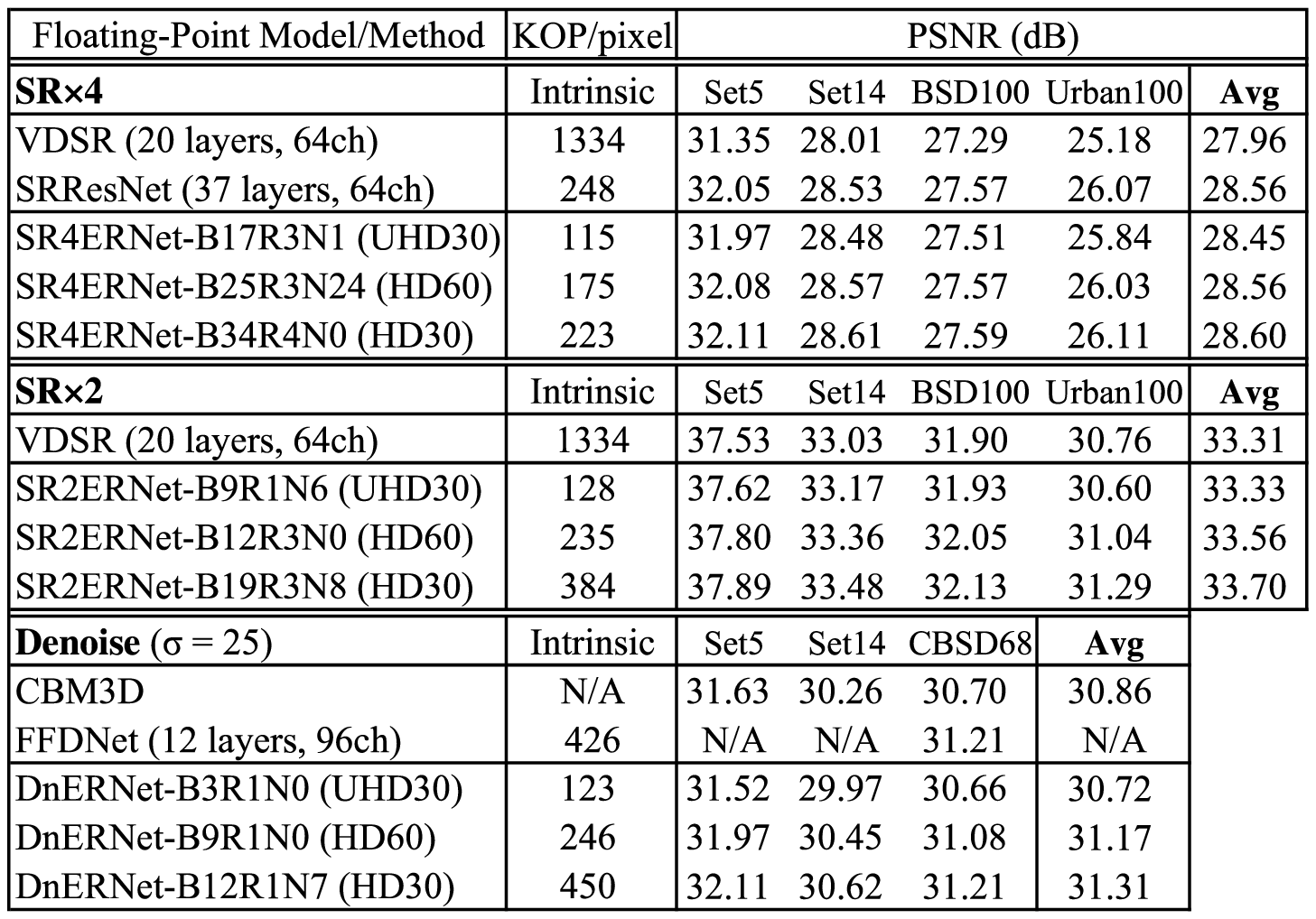} \\
\caption{PSNR performance of polished ERNet models.
}
\label{tab:tab_model_optimization}
\end{table}


\textbf{Fixed-point precision and entropy coding.}
We tested both the L1-norm and L2-norm quantization on the polished models, and the results are shown in Table \ref{tab:tab_model_quantization}.
For the case of SR4ERNet for HD30, the bitstreams will exceed the capacity of the parameter memory in Table \ref{tab:tab_eCNN_spec} using 8-bit precision.
Therefore, we further perform 7-bit quantization on some selected parameter groups to match the capacity.
In general, the L1-norm causes more quality degradation at first because more large values are cropped, but it can be well recovered after fine-tuning.
We chose to use the L1-optimized models for their better PSNR quality despite their higher entropy for larger dynamic range, and the compression ratio is around 1.1-1.5$\times$.
As a result, the PSNR drops are well limited between 0.05 to 0.14 dB for using dynamic 8-bit precision and 1,288KB of parameter memory.
In addition, the values of cross entropy are close to the Shannon limits, which justifies the usage of the simple encoding method.

\begin{figure}[t]
\centering
   \includegraphics[width=8.3cm]{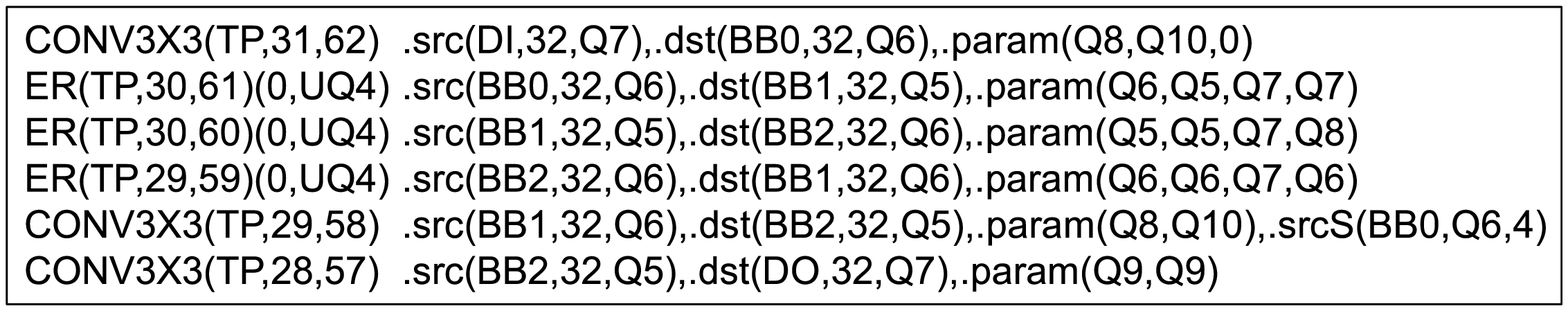} \\
\caption{Program of DnERNet-B3R1N0 (UHD30).
}
\label{fig:fig_Dn_UHD60_program}
\end{figure}

\textbf{Program.}
The coarse-grained FBISA instructions result in concise programs.
Fig. \ref{fig:fig_Dn_UHD60_program} shows a six-line program for the six-layer DnERNet for UHD30.
The attributes of the opcodes specify the sizes of output blocks in terms of 4$\times$2-tiles, and most of other fields identify the dynamic Q-formats.

\begin{table}
\centering
   \includegraphics[width=8.3cm]{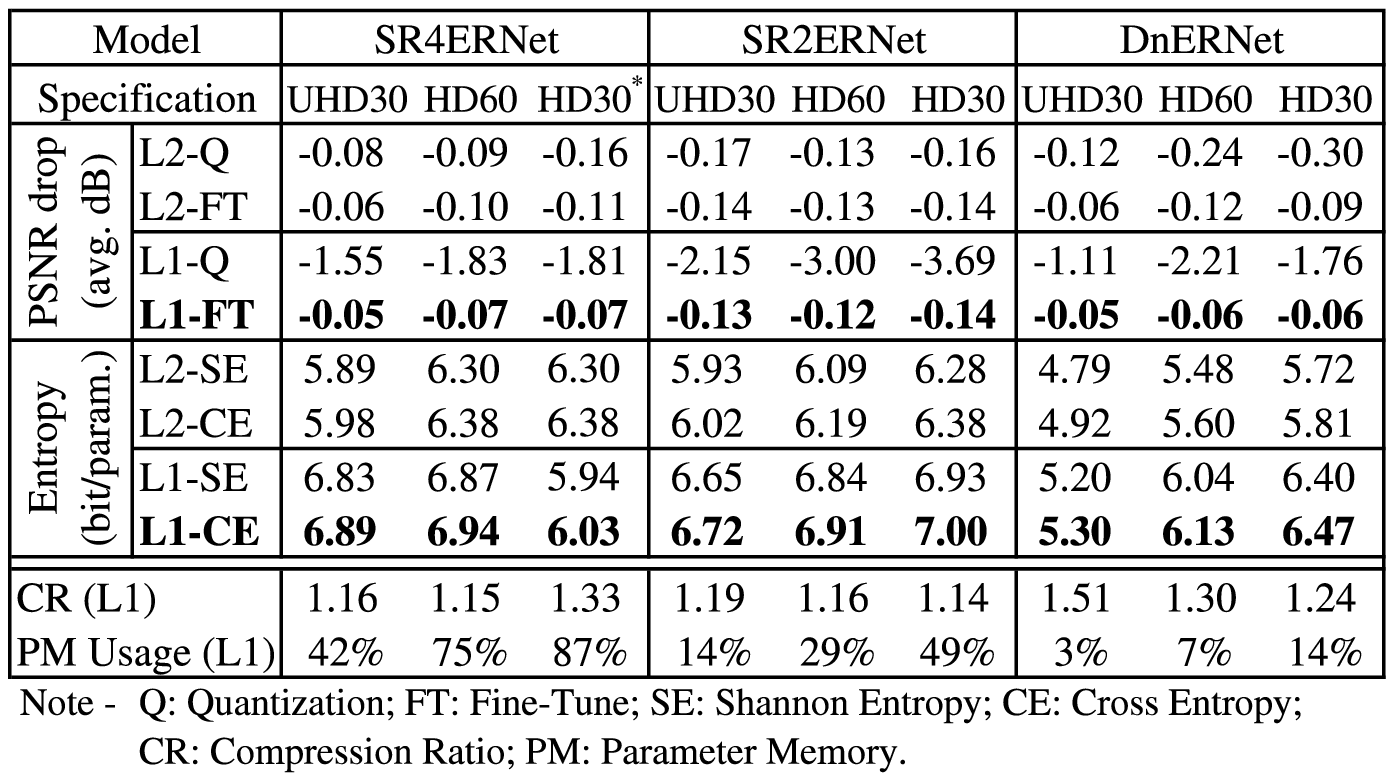} \\
\caption{Model quantization and entropy coding.
}
\label{tab:tab_model_quantization}
\end{table}

\subsection{eCNN Performance}
\label{ssec:eCNNperf}

\textbf{Implementation.}
We implemented eCNN in Verilog HDL for TSMC 40nm technology and used ARM memory compilers to generate all SRAM macros.
We used Synopsys IC Compiler for placement and routing.
We performed layouts for five essential and well-pipelined macro circuits which constitute the eCNN in a collectively exhaustive way.
For fast and accurate power estimation, we ran RTL simulation to generate signal activity waveforms and then propagated them to post-layout netlists with extracted parasitics.

%

\begin{figure}[t]
\centering
   \includegraphics[width=8.3cm]{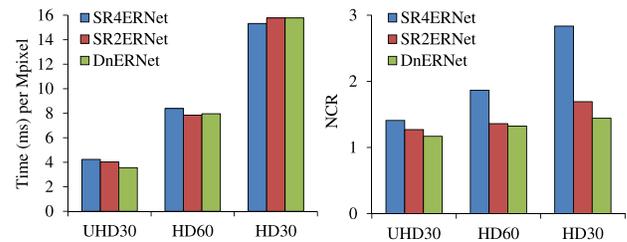} \\
\caption{Inference time (left) and NCR (right).
}
\label{fig:fig_ncr_time}
\end{figure}

\textbf{Layout performance.}
The eCNN processor can run at 250MHz and achieves up to 41~TOPS of inference performance.
The total area is 55.23$\mbox{mm}^2$ and the average power consumption is 6.94W at 0.9V.
The details are summarized in Table \ref{tab:tab_area_power}.
The LCONV3$\times$3 engine delivers 90\% of inference performance and thus occupies the most resource, i.e.~65.8\% of area and 87.4\% of power.
And the LCONV1$\times$1 engine is responsible for the rest 10\% of performance and uses another 7.0\% of area and 6.6\% of power.
On the other hand, the three block buffers (1536KB) and the parameter memory (1288KB) contributes 11.3\% and 7.9\% of area for storing feature maps and parameters respectively.
But they consume only 3.9\% of power in total thanks to well-constrained word depths and highly-optimized SRAM macros.
The computation for ERNets is profiled in Fig. \ref{fig:fig_ncr_time} where the inference time indicates real-time capability and NCR shows computing overheads.
Note that there is a tradeoff between inference time and image quality as shown in Table \ref{tab:tab_model_optimization}.

\begin{table}
\centering
   \includegraphics[width=8.3cm]{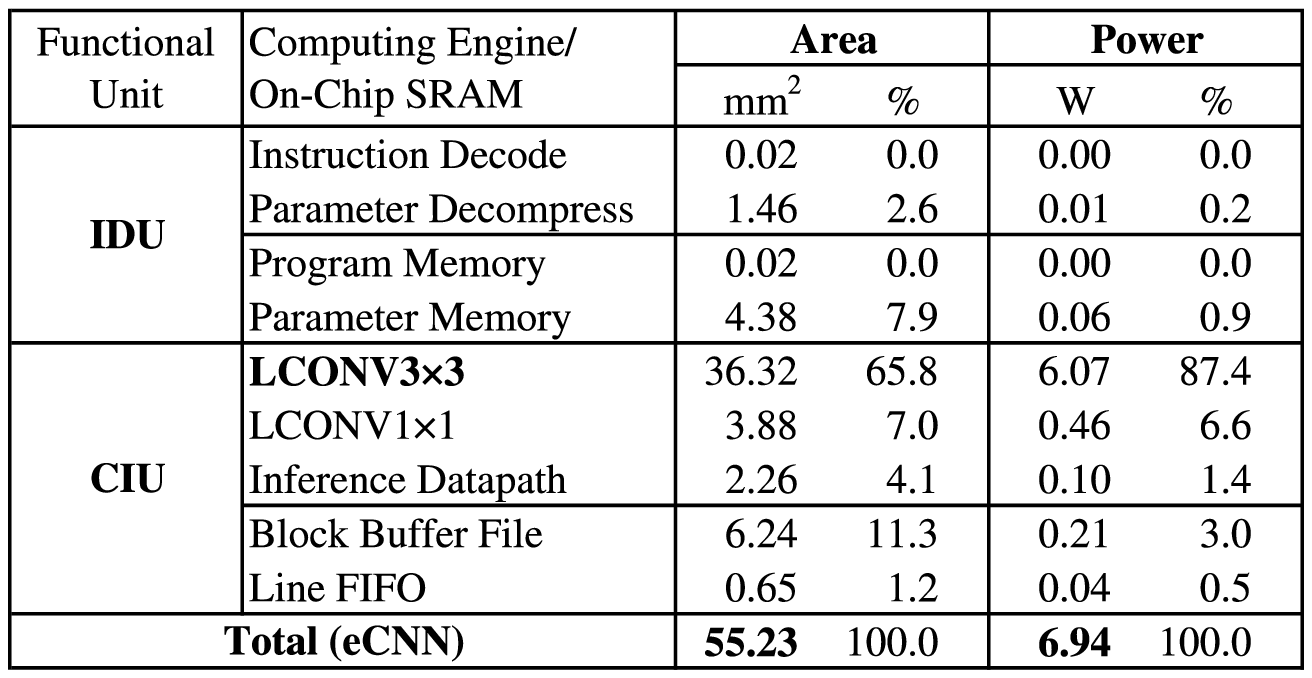} \\
\caption{Area and power consumption of eCNN.
}
\label{tab:tab_area_power}
\end{table}

\begin{figure}[t]
\centering
   \includegraphics[width=8.3cm]{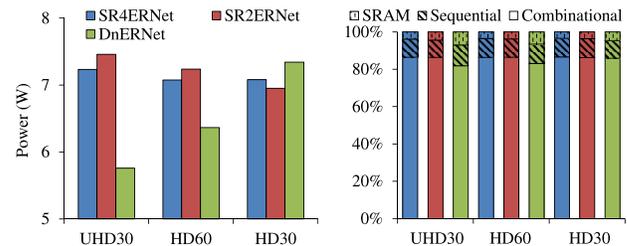} \\
\caption{Power consumption breakdown on ERNet models (left) and circuit types (right).
}
\label{fig:fig_power}
\end{figure}

\textbf{Power breakdown.}
The power consumption for each ERNet model is shown in Fig. \ref{fig:fig_power}.
We found that the variation between different specifications is related to the quality difference.
For example, DnERNets have the largest power variation, 1.58W, while they have 0.58 dB of PSNR drop from HD30 to UHD30.
In contrast, SR4ERNets have the smallest variations in both power consumption and PSNR.
Fig. \ref{fig:fig_power} also shows the breakdown for three circuit types.
The combinational circuits contribute 82-87\% of power consumption for the highly-parallel convolution.
The sequential circuits constantly occupy about 10\% for the locally-distributed parameter registers, 4$\times$2-tile pipeline registers, and clock tree.
The rest 3-7\% is then consumed by SRAMs.

\begin{figure}[t]
\centering
   \includegraphics[width=8.3cm]{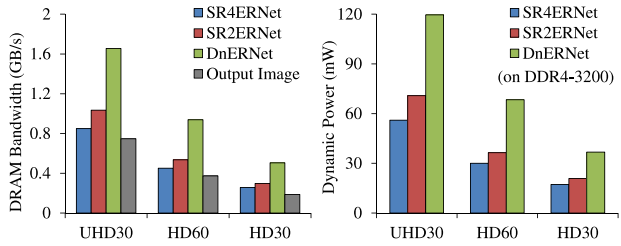} \\
\caption{DRAM bandwidth and dynamic power.
}
\label{fig:fig_dram}
\end{figure}

\textbf{DRAM bandwidth and power.}
The DRAM access via the data input and output FIFOs is highly regular and can be optimized in a deterministic way.
The DRAM bandwidth and dynamic power consumption for each ERNet model are shown in Fig. \ref{fig:fig_dram}.
The DnERNets require the most bandwidth for each specification: 1.66GB/s for UHD30, 0.94GB/s for HD60, and 0.5GB/s for HD30.
But the NBRs are still only 2.2$\times$, 2.5$\times$, and 2.7$\times$, respectively.
Therefore, the eCNN can support high-end applications with low-end DRAM configurations.
For example, DDR-400 (3.2GB/s), DDR-266 (2.1GB/s), and DDR-200 (1.6GB/s) are sufficient for UHD30, HD60, and HD30 respectively.
Regarding power consumption, we use Micron DDR4 SDRAM System-Power Calculator \cite{micron} for evaluation on DDR4-3200.
The small bandwidth of eCNN consumes only less than 120mW of dynamic power (activation/read/write) while the leakage power consumes 267mW.

\begin{table*}
\centering
   \includegraphics[width=17.4cm]{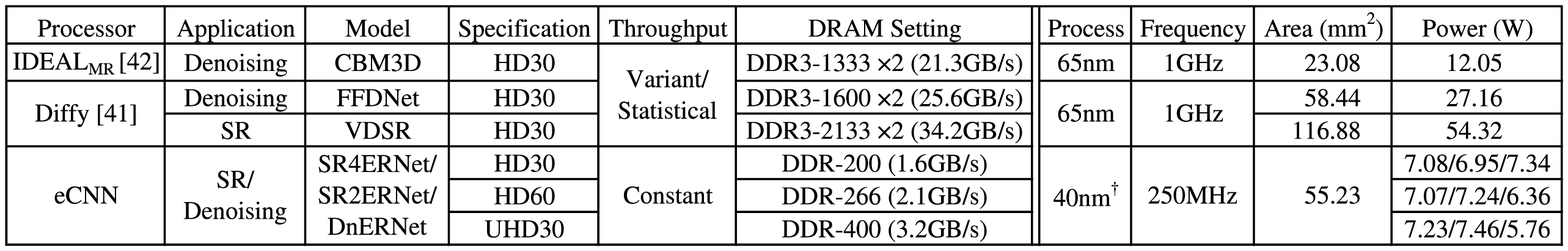} \\
   \hfill \scriptsize{$\dagger$ The 40nm technology outperforms its 65nm counterpart at half of the power consumption under the same operation speed \cite{tsmc_40nm}. $\;\;$}
\caption{Comparison of computational imaging processors.
}
\label{tab:tab_comparison}
\end{table*}

\textbf{Comparison.}
Table \ref{tab:tab_comparison} compares the eCNN with two state-of-the-art processors for computational imaging applications: IDEAL \cite{ideal_2017} for BM3D and Diffy \cite{diffy_2018} for CNN.
Both of them can only support the HD30 specification and already require high-end DRAM settings, i.e.~dual-channel DDR3-1333 to DDR3-2133.
However, eCNN can deliver up to UHD30 performance using only DDR-400.
Another advantage of eCNN is its constant pixel throughput to facilitate real-time applications.
In contrast, the performance of IDEAL and Diffy highly varies with input images since statistical properties are deployed for acceleration.

To compare power consumption, we list the reported numbers from \cite{ideal_2017} and \cite{diffy_2018} for IDEAL and Diffy on the right of Table \ref{tab:tab_comparison}.
Note that these numbers cannot be used to determine superiority directly because they are highly related to technology nodes, implementation details, and deployed models/algorithms.
For denoising at HD30, IDEAL needs 12.05W for BM3D and Diffy demands 27.16W (8 tiles) for FFDNet; however, eCNN consumes only 7.34W for DnERNet which is 0.39~dB better than CBM3D and comparable to FFDNet.
For four-times SR at HD30, Diffy demands 54.32W (16 tiles) for VDSR while eCNN consumes only 7.08W for SR4ERNet (0.57~dB better than VDSR).

We also used a CNN accelerator simulator, SCALE-Sim \cite{scale_sim}, to simulate the performance of ERNets with the same processor configuration as
the classical TPU \cite{tpu_2017}.
Note that TPU is a high-performance 28nm processor which provides 92~TOPS at 40~W and has 28~MB of SRAM to store feature maps and parameters for data reuse.
The simulation shows that 4K UHD 21.9 fps and Full HD 55.3 fps are achieved for SR4ERNet-B17R3N1 and SR4ERNet-B34R4N0 respectively.
And the required DRAM bandwidths are 12.2 GB/s and 8.3 GB/s.
As a result, eCNN provides 3.1$\times$ and 1.2$\times$ of throughput efficiency (fps/TOPS) and, in particular, 6.4$\times$ and 14.4$\times$ of arithmetic intensity (TOPS/GB/s), respectively, for these two models.
This also demonstrates the advantage of our joint-design approach for computational imaging tasks.

\subsection{Computer Vision Applications}
\label{ssec:otherexample}

\begin{figure}
\centering
   \includegraphics[width=8.2cm]{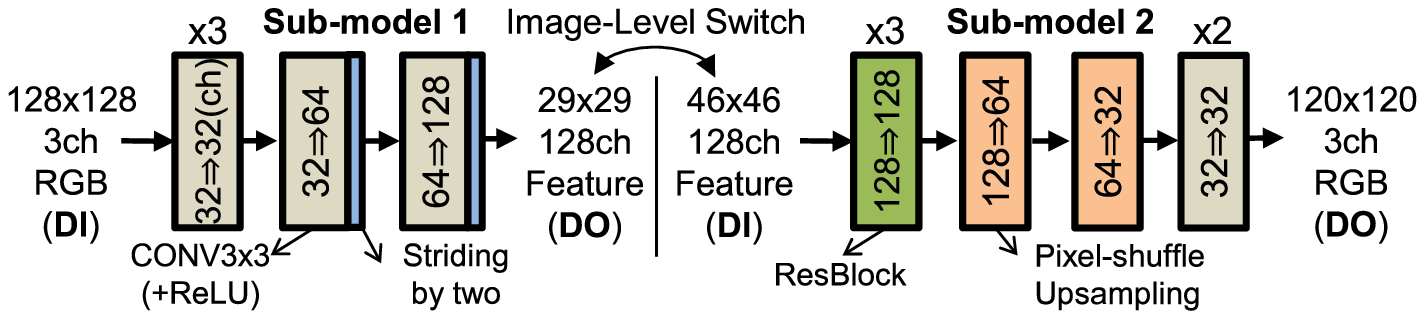} \\
   (a)\\
   \includegraphics[width=8.3cm]{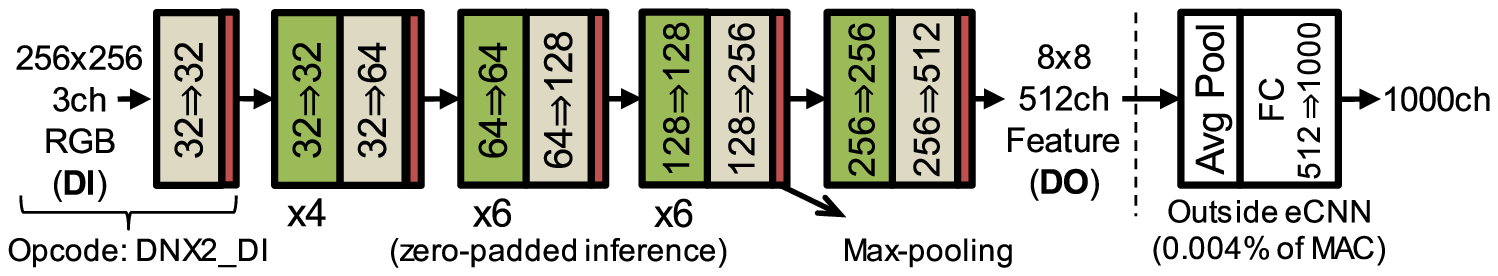} \\
   (b)
\caption{Computer vision models based on FBISA: (a) style transfer and (b) object recognition.
}
\label{fig:fig_other_example}
\end{figure}

\textbf{Model structure.}
To show the model flexibility of our approach, we built FBISA-compatible models for style transfer and object recognition as shown in Fig. \ref{fig:fig_other_example}.
They differ from the ones for computational imaging mainly in three respects: spatial downsampling, wider channels, and batch normalization layers.
The first two are supported in FBISA by concatenating 32ch leaf-modules.
The last one is used to stabilize model pre-training and will be merged into convolutional layers for quantization and inference.

\textbf{Style transfer.}
We used two downsamplers to increase the receptive field as suggested in \cite{st_sr_2016}.
Since this will increase NCR significantly, we split the model into two sub-models as shown in Fig. \ref{fig:fig_other_example}(a) to reduce computing overheads.
After training and quantization, our model can deliver similar style transfer effects with \cite{st_sr_2016}.
The performance on eCNN is Full HD 29.5 fps for this model while Nvidia Titan X GPU is used in \cite{st_sr_2016} and only generates 512$\times$512 20fps.
In addition, the DRAM bandwidth of our approach is only 1.91GB/s, which can enable this advanced application on embedded devices.

\textbf{Object recognition.}
We devised a 40-layer residual network in Fig. \ref{fig:fig_other_example}(b) for eCNN to perform object recognition.
To reduce the amount of parameters, we avoided 512ch ResBlocks and instead put more computation in thinner layers.
The final 8-bit model achieves 69.7\% top-1 accuracy for ImageNet \cite{imagenet} with 5M parameters.
The performance is comparable to ResNet-18 (69.6\%; 11M) and VGG-16 (71.5\%; 138M) \cite{vggnet_2015}.
To support this model, we need to increase the size of parameter memory by three times, 
and the area of eCNN would become 63.99 $\mbox{mm}^2$.
Then the performance achieves 1344 fps (0.74 ms per image) with 308 MB/s of DRAM bandwidth.
For each image, eCNN only consumes 5.25 mJ of energy and 231 KB of DRAM access.
For comparison, the Eyeriss \cite{eyeriss_jssc_2016}, which has 12.25 $\mbox{mm}^2$ of core area with 65nm technology, delivers 0.7 fps (4.3 s for a batch of three images) with 236mW of power consumption and 74 MB/s of DRAM bandwidth for VGG-16.
Thus it demands as high as 337 mJ of energy and 106 MB of DRAM access for one image.
In this case study, we demonstrate the model flexibility of eCNN and also show that our joint hardware-model approach can benefit object recognition tasks as well.

\section{Related Work}
\label{sec:relatedwork}

\textbf{Instruction set.}
Previous SIMD works usually devised load instructions for parameters and adopted medium-grained operands for features, such as vector/matrix \cite{cambricon_2016}, 2D tile \cite{deephi_2016}, and compute tile \cite{computileISA_2017}, for providing flexibility.
In contrast, we apply a parameter-inside approach and large-grained feature operands for optimizing power consumption and computing capability for highly-parallel convolution.

\textbf{Model structure.}
Most of previous hardware-oriented models aim to reduce complexity.
In particular, SqueezeNet \cite{squeezenet_2017} temporarily reduces model width and then expands back for residual connections.
And MobileNetV2 \cite{mobilenetv2_2018} moves the connections to thinner layers to reduce storage.
This results in an expansion-reduction structure similar to ERNet.
However, our goal is to increase complexity under hardware constraints, and thus the implementation details are quite different.

\textbf{Winograd convolution.}
It is an efficient algorithm to reduce multipliers for CONV3$\times$3 and recently shows advantages on GPU \cite{winograd_2016}, FPGA \cite{winograd_fpga_dac2018}, and embedded processor \cite{winograd_myriad_dac2018}.
However, it increases 23.5\% of area in our case because the overheads of long internal bitwidths and additional pre-/post-processing become significant for our 8-bit implementation.
Therefore, we used a direct implementation instead.

\textbf{Cross-frame optimization.}
It is a new research direction to reduce CNN computation by exploiting temporal redundancy \cite{Euphrates_2018,Eva2_2018} or input similarity \cite{reuse_2018} across video or audio frames.
Moreover, this concept has also been applied to compensate the unreliability brought by pruning \cite{darkside_2018}.
This direction is complementary to our approach and can be used to further enhance the performance of eCNN.

\section{Conclusion}
\label{sec:conclusion}

In this paper, we investigate a hardware-first framework to support computational imaging CNNs for up to 4K Ultra-HD applications on edge devices.
Instead of accelerating existing models, we devise the hardware-oriented ERNets for the adopted block-based inference flow which can eliminate DRAM bandwidth for feature maps.
For providing high computing capability, we construct the coarse-grained FBISA to enable highly-parallel convolution.
Finally, we implement the high-performance eCNN processor to incorporate ERNet and FBISA.
The training and layout results show that this framework can provide superior hardware performance and image quality at the same time.
In addition, its flexibility is demonstrated by the usage examples of style transfer and object recognition.
Our future work is to include more CNN variants for different applications and unleash their power on edge devices.

\section*{Acknowledgments}

This work was supported by the Ministry of Science and Technology, Taiwan, R.O.C. under Grant no. MOST 107-2218-E-007-029.




\appendix
\renewcommand\thefigure{\thesection.\arabic{figure}}  
\setcounter{figure}{0}   
\renewcommand\thetable{\thesection.\arabic{table}}  
\setcounter{table}{0}  
\section{Model Variants for Denoising} 

\begin{figure}[htb!]
\centering
   \includegraphics[width=8.3cm]{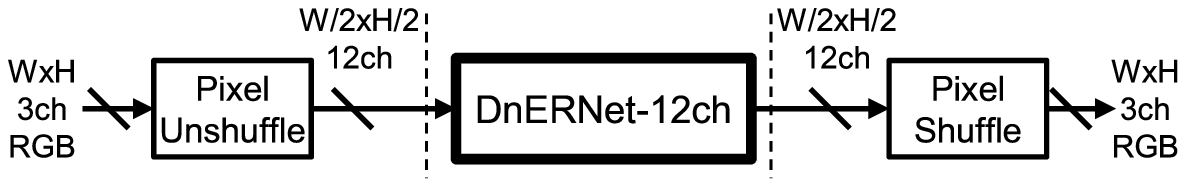} \\
\caption{DnERNet-12ch for denoising.
}
\label{fig:fig_dnernet12ch}
\end{figure}

The image quality of DnERNets drops quickly for higher specifications due to shallower layers.
Here, under the eCNN framework, we show how to improve their quality using the same downsampling strategy of FFDNet.
As shown in Fig. \ref{fig:fig_dnernet12ch}, we can perform pixel unshuffle, which packs 2$\times$2 3-ch RGB pixels into a 12-ch one, for the input image.
Then we construct DnERNet-12ch models as DNERNets, and the only difference is their input and output channels both become twelve, instead of three.
Finally, the output image can be obtained by performing pixel shuffle.

The PSNR performance of the picked and then polished models is show in Table \ref{tab:tab_model_optimization-12ch}.
For the UHD30 specification, the DnERNet-12ch-B8R2N5 outperforms the DnERNet-B3R1N0 by 0.54 dB and now provides FFDNet-level quality.
On the other hand, the DnERNet-12ch-B19R3N15 for HD30 even surpasses FFDNet by 0.15 dB using smaller intrinsic complexity.
Lastly, the quantized 8-bit models have 0.09 dB drop on average, which are similar to ordinary ERNets, and the DRAM bandwidth is only at most 1.8GB/s.

\begin{table}[htb!]
\centering
   \includegraphics[width=8.0cm]{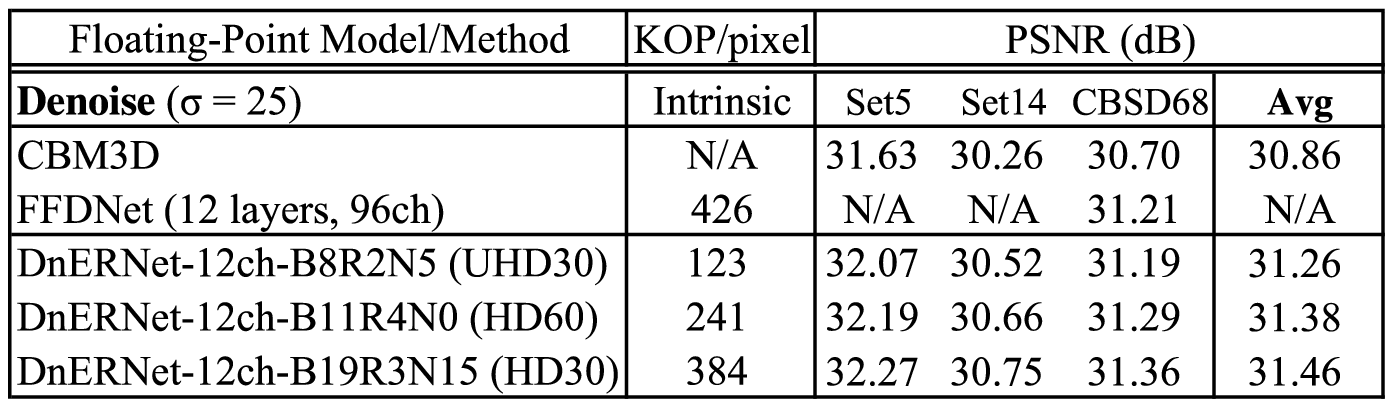} \\
\caption{PSNR of polished DnERNet-12ch models.
}
\label{tab:tab_model_optimization-12ch}
\end{table}

\end{document}